\documentstyle[12pt,epsf,rotate,epic]{article} 
\if@twoside \oddsidemargin 21pt \evensidemargin 59pt \marginparwidth 85pt
\else \oddsidemargin 0pt \evensidemargin 0pt
\fi
\newcounter{saveeqn}                  
\newcommand{\alpheqn}[1]{\refstepcounter{equation}\label{#1}%
\setcounter{saveeqn}{\value{equation}}%
\setcounter{equation}{0}%
\renewcommand{\theequation}
{\mbox{\arabic{saveeqn}\alph{equation}}}}
\newcommand{\reseteqn}{\setcounter{equation}{\value{saveeqn}}%
\renewcommand{\theequation}{\arabic{equation}}}
\newcounter{savefig}
\newcommand{\alphfig}[1]{\refstepcounter{figure}\label{#1}%
\setcounter{savefig}{\value{figure}}%
\setcounter{figure}{0}%
\renewcommand{\thefigure}
{\mbox{\arabic{savefig}\alph{figure}}}}
\newcommand{\resetfig}{\setcounter{figure}{\value{savefig}}%
\renewcommand{\thefigure}{\arabic{figure}}}
\newcommand{\capt}[1]{\newcommand{\groesse}{\normalsize}%
\renewcommand{\normalsize}{\footnotesize}%
\caption[ ]{#1}%
\renewcommand{\normalsize}{\groesse}}

\renewcommand{\bibitem}[1]{\\[0.2cm]\vphantom{#1}}
\renewcommand{\vec}[1]{\mbox{\boldmath $#1$}}
\newcommand{\weg}[1]{#1}                       
\begin{document}
\vfill
\begin{center}
{\large Dirk Helbing}\\[0.7cm] 
{\Large A stochastic behavioral model and a 
`microscopic' foundation of evolutionary game theory}\\[0.8cm]
\end{center}
{Running head: Stochastic game theory}
\begin{abstract}
A stochastic model for be\-ha\-vio\-ral chan\-ges by 
imitative pair interactions of individuals is developed. 
`Microscopic' assumptions on the specific form of the imitative
processes lead to a stochastic version of the game dynamical
equations. That means, the approximate mean value equations of 
these equations are the game dynamical equations
of evolutionary game theory. 
\par
The stochastic version of the game dynamical equations allows
the derivation of covariance equations. These should always be solved 
along with the ordinary game dynamical equations.
On the one hand, the average behavior is affected by the
covariances so that the game dynamical equations must be corrected
for increasing covariances. Otherwise they may become invalid in the course
of time. On the other hand, the covariances are
a measure for the reliability of game dynamical
descriptions. An increase of the covariances beyond a critical value
indicates a phase transition, i.e. a sudden change
in the properties of the considered social system.
\par
The applicability and use of the introduced equations are illustrated by
computational results for the social self-organization of behavioral 
conventions.
\end{abstract}
{Keywords: evolutionary game theory, behavioral model, imitative processes,
self-or\-gan\-ization of behavioral conventions, stochastic game theory, 
mean value equations, covariance equations, reliability of rate
equations, expected strategy distribution, most probable
strategy distribution}
\vfill
\clearpage

\section{Introduction}

This paper treats a mathematical model 
for the temporal change of the proportions
of individuals showing certain behavioral strategies. Models of this
kind are of special interest for a {\em quantitative understanding}
or {\em prognosis} of social developments. For the description of the
{\em competition} or {\em cooperation} in populations there already exist
{\em game theoretical approaches} (cf. e.g. 
{\sc von Neumann} and {\sc Morgenstern},
1944; {\sc Luce} and {\sc Raiffa}, 1957; 
{\sc Rapoport} and {\sc Chammah}, 1965; {\sc Axelrod}, 1984). 
In order to cope with time-dependent problems the method of
{\em iterated games} has been developed and
used for a long time. However, some years
ago, the {\em game dynamical equations} have been discovered
({\sc Taylor} and {\sc Jonker}, 1978; {\sc Hofbauer} et. al., 1979; 
{\sc Zeeman}, 1980). These
are ordinary differential equations, which are related to the theory of
evolution ({\sc Eigen}, 1971; {\sc Fisher}, 1930; {\sc Eigen} and
{\sc Schuster}, 1979; {\sc Hofbauer} and {\sc Sigmund}, 1988; {\sc Feistel}
and {\sc Ebeling}, 1989). 
Therefore, one also speaks of {\em evolutionary
game theory}. 
\par
The game dynamical equations have the
following advantages:
\begin{itemize}
\item They are continuous in time which is more adequate for many
problems.
\item Ordinary differential equations are easier to handle than
iterated formulations.
\item Analytical results can be derived more easily
(cf. e.g. {\sc Hofbauer} and {\sc Sigmund}, 1988;
{\sc Helbing}, 1992).
\end{itemize}
Up to now, there only exists a {\em `macroscopic'} foundation of the game
dynamical equations, i.e. a derivation from a 
{\em collective} level of behavior (cf. Section \ref{Macr}).
In this paper a {\em `microscopic'} foundation will be given, i.e. a
derivation on the basis of the {\em individual} behavior. 
With this aim in view, we will first develop a {\em stochastic} behavioral
model for the following reasons:
\begin{itemize}
\item A stochastic model, i.e. a model that can describe
random {\em fluctuations} of the quantities of interest, 
can cope with the fact that behavioral changes are not
exactly predictable (which is a consequence of the `freedom of 
decision-making').
\item The phenomena appearing in the considered social system can be
connected to the principles of {\em individual} behavior. As a consequence,
processes on the `macroscopic' (collective) level can be understood as
effects of `microscopic' (individual) interactions.
\item The probability of occurence of each strategy can be calculated.
This is especially important for small social systems 
which are subject to large fluctuations (since they consist
of a few individuals only).
\item The stochastic model allows the derivation of {\em covariance equations}
(cf. Section~\ref{Mostexp}). Since the covariances influence the average
temporal behavior, they are an essential criterium for the
validity and reliability of behavioral descriptions
by rate equations (which are actually approximate mean value equations). 
If the covariances exceed a certain critical value,
this indicates the occurence of a {\em phase transition}, i.e.
a sudden change of the properties of the considered social system.
\end{itemize}
For the description of systems that are subject to random fluctuations
different stochastic methods have been developed
(cf. e.g. {\sc Gardiner}, 1983; {\sc Weidlich} and {\sc Haag}, 1983; 
{\sc Helbing}, 1992). One method is to delineate
the temporal evolution of the {\em probability distribution} over the different
possible states (which represent behavioral strategies, here). This method
is particularly suitable for an {\em `ensemble'} of similar systems or
for frequently occuring processes. In the case of {\em discrete} states, the 
{\em master equation} has to be used, whereas in the case of {\em continuous}
state variables the {\sc Fokker-Planck} {\em equation} is normally preferred
since it is easier to handle ({\sc Fokker}, 1914; {\sc Planck}, 1917). 
The {\sc Fokker-Planck} equation can, in good
approximation, also be applied to systems with a large number of discrete
states if state changes only occur between neighbouring states. It 
can be derived from the master equation by a {\sc Kramers-Moyal} {\em
expansion} ({\sc Kramers}, 1940; {\sc Moyal}, 1949),
i.e. a second order {\sc Taylor} {\em approximation}, then.
\par
Another method, the {\sc Langevin} {\em equation} (1908) (or             
{\em stochastic differential equation}) is applied to
the description of the temporal evolution of
{\em single} fluctuation-affected systems. It consists of a deterministic
dynamical part which delineates systematic state changes and a stochastic
fluctuation term which reflects random state variations. The 
{\sc Langevin} equation can be reformulated in terms of a {\sc Fokker-Planck}
equation and vice versa (if the fluctuations are {\sc Gauss}ian and
$\delta$-correlated which is normally the case; 
cf. {\sc Stratonovich}, 1963, 1967; {\sc Weidlich} and {\sc Haag}, 1983).
\par
Although these methods come from statistical physics, the application to
{\em interdisciplinary} topics has meanwhile a long and successful tradition,
beginning with the work of {\sc Weidlich} (1971, 1972), {\sc Haken} (1975),
{\sc Prigogine} (1976), {\sc Nicolis} and {\sc Prigogine} (1977). 
Also for social and economic processes
{\sc Fokker-Planck} equation models
(cf. e.g. {\sc Weidlich} and {\sc Haag}, 1983; {\sc Topol}, 1991) 
as well as master equation models (cf. e.g. {\sc Weidlich} and {\sc Haag},
1983; {\sc Weidlich}, 1991; {\sc Haag} et.al., 1993; 
{\sc Weidlich} and {\sc Braun}, 1992) were proposed. 
In this paper we will develop a
behavioral model on the basis of the master equation (Section \ref{stoch}). 
For this purpose we have to specify the {\em transition rates}, i.e. the
{\em probabilities} per time unit with which changes of behavioral strategies
take place. The transition rates can be decomposed into 
\begin{itemize}
\item rates describing {\em spontaneous} strategy changes, and
\item rates describing strategy changes due to {\em pair interactions}
of individuals.
\end{itemize}
In the following we will restrict our considerations
to {\em imitative} pair interactions which
seem to be the most important ones ({\sc Helbing}, 1994).
By distinguishing several {\em subpopulations} $a$, 
different {\em types} of behavior or different {\em groups} of individuals
can be taken into account.
\par
In order to connect the stochastic behavioral model to the game dynamical
equations
the transition rates have to be chosen in such a way that they depend on the
{\em expected successes} of the behavioral strategies (cf. Section
\ref{Micfound}). 
The ordinary game dynamical equations are
the {\em approximate mean value} equations of the stochastic behavioral
model (cf. Section \ref{Deriv}).
\par
For the approximate mean value equations correction terms can be calculated. 
These depend on the covariances 
(of the numbers of individuals pursuing a certain strategy) (cf. Section
\ref{Corr}). Neglecting these corrections, the game dynamical equations
may lose their validity after some time. The calculation
of the covariances allows the determination of 
the time interval during which game
dynamical descriptions are reliable (cf. Section \ref{CompRes}). 
\par
The introduced equations are illustrated by computational results for
the self-organization of a behavioral convention by a
competition between two alternative, but equivalent strategies 
(cf. Sections \ref{Selfor}
and \ref{Mostexp}). These results are relevant for economics
with respect to the rivalry between similar products
({\sc Arthur}, 1988, 1989; {\sc Hauk}, 1994).

\section{The stochastic behavioral model} \label{stoch}

Suppose we consider a social system with $N$ individuals. These individuals
can be divided into $A$ {\em subpopulations} $a$ consisting of 
$N_a$ individuals, i.e.
\begin{displaymath}
 \sum_{a=1}^A N_a = N \, .
\end{displaymath}
By subpopulations different social groups (e.g. blue and white 
collars) or different characteristic {\em types of behavior} are
distinguished. In the following we will assume that individuals of the same 
subpopulation (group) behave {\em cooperatively} 
due to common interests, whereas individuals
of different subpopulations (groups)
do not so due to {\em conflicting} interests.
\par
The $N_a$ individuals of each subpopulation $a$ are distributed
over several {\em states}
\begin{displaymath}
 i \in \{1,\dots,S\} 
\end{displaymath}
which represent the alternative {\em (behavioral)
strategies} of an individual. For the time being, every individual shall be
able to choose each of the $S$ strategies, i.e. the same strategy set shall be
available for each subpopulation.
If the {\em occupation number} $n_i^a(t)$
denotes the number of individuals of subpopulation $a$ who use 
strategy $i$ at the time $t$, we have the relation
\begin{equation}
 \sum_{i=1}^S n_i^a(t) = N_a \, .
\label{sum}
\end{equation}
Let 
\begin{displaymath}
 \vec{n} := (n_1^1,\dots,n_i^a,\dots,n_S^A)
\end{displaymath}
be the vector consisting of all occupation numbers $n_i^a$. This vector is
called the {\em socioconfiguration} since it contains all information
about the distribution of the $N$ individuals over the states $i$. 
$P(\vec{n},t)$ shall denote the {\em probability} to find the
socioconfiguration $\vec{n}$ at the time $t$. This implies
\begin{displaymath}
 0 \le P(\vec{n},t) \le 1 \qquad \mbox{and} \qquad
\sum_{\weg{n}} P(\vec{n},t)= 1 \, .
\end{displaymath}
If transitions
from socioconfiguration $\vec{n}$ to $\vec{n}'$
occur with a probability of $P(\vec{n}',t+\Delta t|\vec{n},t)$ 
during a short time interval
$\Delta t$, we have a {\em (relative) transition rate} of
\begin{displaymath}
 w(\vec{n}'|\vec{n};t) 
:= \lim_{\Delta t \rightarrow 0}
\frac{P(\vec{n}',t+\Delta t|\vec{n},t)}{\Delta t} \, .
\end{displaymath}
The {\em absolute} transition rate of changes
from $\vec{n}$ to $\vec{n}'$ is the product 
$w(\vec{n}'|\vec{n};t)P(\vec{n},t)$ of the probability
$P(\vec{n},t)$ to have configuration $\vec{n}$
and the {\em relative} transition rate $w(\vec{n}'|\vec{n};t)$
if having configuration $\vec{n}$. Whereas the {\em inflow} into
$\vec{n}$ is given as
the sum over all absolute transition rates of changes from an {\em arbitrary}
configuration $\vec{n}'$ to $\vec{n}$, the {\em outflow} from $\vec{n}$
is given as the sum over all absolute transition rates of changes
from $\vec{n}$ to {\em another} configuration $\vec{n}'$. Since the
temporal change of the probability $P(\vec{n},t)$ is determined
by the inflow into $\vec{n}$ reduced by the outflow from $\vec{n}$, we
find the socalled {\em master equation}             
\begin{eqnarray}
 \frac{d}{dt} P(\vec{n},t) &=&
\mbox{inflow into $\vec{n}$ } - \mbox{ outflow from $\vec{n}$} \nonumber \\
&=& \sum_{\weg{n}'} w(\vec{n}|\vec{n}';t)P(\vec{n}',t) 
- \sum_{\weg{n}'} w(\vec{n}'|\vec{n};t)P(\vec{n},t) \qquad
\label{master}
\end{eqnarray}
({\sc Pauli}, 1928; {\sc Haken}, 1979; {\sc Weidlich} and {\sc Haag},
1983; {\sc Weidlich}, 1991). 
\par
It will be assumed that two processes contribute to a change of the
socioconfiguration $\vec{n}$:
\begin{itemize}
\item Individuals may change their strategy $i$ spontaneously and
independently of each other to another strategy $i'$ with an
{\em individual} transition rate $\widehat{w}_a(i'|i;t)$. 
These changes correspond to transitions of the socioconfiguration from
$\vec{n}$ to                                                                  
\[
 \vec{n}_{i'i}^a := (n_1^1,\dots,(n_{i'}^a + 1), 
 \dots, (n_i^a -1), \dots,n_S^A)                                
\]
with a {\em configurational} transition rate $w(\vec{n}_{i'i}^a|\vec{n};t)=
n_i^a \widehat{w}_a(i'|i;t)$ which is
proportional to the number $n_i^a$ of individuals who can change 
strategy $i$.
\item An individual of subpopulation $a$ may change the strategy from
$i$ to $i'$ during a pair interaction with 
an individual of some subpopulation
$b$ who changes the strategy from $j$ to $j'$. Let transitions of this kind
occur with a probability $\widehat{w}_{ab}(i',j'|i,j;t)$ per time unit. The
corresponding change of the socioconfiguration from $\vec{n}$ to
\[
 \vec{n}_{i'j'ij}^{ab} 
:= (n_1^1,\dots,(n_{i'}^a + 1),\dots, 
(n_i^a -1),\dots,(n_{j'}^b +1), 
\dots,(n_j^b-1),\dots,n_S^A)
\]
leads to a configurational transition rate
$w(\vec{n}_{i'j'ij}^{ab}|\vec{n};t)
= n_i^a n_j^b \widehat{w}_{ab}(i',j'|i,j;t)$
which is proportional to the number
$n_i^an_j^b$ of possible pair interactions between individuals of
subpopulations $a$ and $b$ who pursue strategy $i$ 
and $j$ respectively.
(Exactly speaking---in order to exclude 
self-interactions---\mbox{$n_i^a n_i^a\widehat{w}_{aa}(i',j'|i,i;t)$} 
has to be replaced by
\mbox{$n_i^a(n_i^a-1)\widehat{w}_{aa}(i',j'|i,i;t)$} if 
$\sum_{j'}\widehat{w}_{aa}(i',j'|i,i;t)
\ll \widehat{w}_a(i'|i;t)$ is invalid and $P(\vec{n},t)$ is
not negligible where $n_i^a \gg 1$ is not fulfilled.)
\end{itemize}
The resulting configurational transition rate $w(\vec{n}'|\vec{n};t)$ is
given by
\begin{equation}
w(\vec{n}'|\vec{n};t) 
:= \left\{
\begin{array}{ll}
n_i^a \widehat{w}_a(i'|i;t) & \mbox{if } \vec{n}' = \vec{n}_{i'i}^a \\
n_i^a n_j^b \widehat{w}_{ab}(i',j'|i,j;t) & \mbox{if }
\vec{n}' = \vec{n}_{i'j'ij}^{ab} \\
0 & \mbox{otherwise.}
\end{array}\right. 
\label{rate}
\end{equation}
As a consequence, the explicit form of master equation (\ref{master})
is
\begin{eqnarray}
\frac{d}{dt}P(\vec{n},t) 
&=& \sum_{a,i,i'}
\Big[ (n_{i'}^a+1)\widehat{w}_a(i|i';t)P(\vec{n}_{i'i}^a,t) 
 - n_i^a \widehat{w}_a(i'|i;t)P(\vec{n},t) \Big] \nonumber \\
&+& \frac{1}{2} \sum_{a,i,i'} \sum_{b,j,j'} \Big[
(n_{i'}^a+1)(n_{j'}^b+1)
\widehat{w}_{ab}(i,j|i',j';t)P(\vec{n}_{i'j'ij}^{ab},t) \nonumber\\
& & \qquad \qquad - \; 
n_i^a n_j^b \widehat{w}_{ab}(i',j'|i,j;t)P(\vec{n},t) \Big] 
\label{MAster}
\end{eqnarray}                          
(cf. {\sc Helbing}, 1992a).
\par
We restricted our considerations to pair interactions, here, since they
normally play the most significant role. Even in groups the most frequent
interactions are alternating pair interactions---not always but in many cases.
In situations where simultaneous interactions between more than two 
individuals are essential (one example for this is {\em group pressure}), 
the above master equation must be extended by higher order interaction terms. 
The corresponding procedure is
discussed by {\sc Helbing} (1992, 1992a).

\section{Stochastic version of the game dynamical equations} \label{Stogame}

\subsection{Specification of the transition rates}

The pair interactions
\begin{equation}
 i',j' \longleftarrow i,j
\label{orgform}
\end{equation}
of two individuals of subpopulations $a$ and $b$ who change their
strategy from $i$ and $j$ to $i'$ 
and $j'$ respectively can be classified into three different
{\em kinds} of processes:
Imitative processes, avoidance processes, and compromising processes. These
are discussed in detail and simulated in several publications  
({\sc Helbing}, 1992, 1992b, 1994). In the following we will focus to 
{\em imitative processes} (processes of persuasion) which describe the tendency
to take over the strategy of another individual. These are of the special 
form\alpheqn{transform}
\begin{eqnarray}
i,i \longleftarrow  i,j & (i\ne j) \, ,\\
j,j \longleftarrow i,j & (i\ne j) \, .
\end{eqnarray}\reseteqn
The corresponding pair interaction rates read\alpheqn{specform}
\begin{eqnarray}
 \widehat{w}_{ab}(i',j'|i,j;t) 
 &=& \widehat{\nu}_{ab} p_{ba}(i|j;t)
  \delta_{ii'} \delta_{ij'} (1-\delta_{ij}) \\
 &+& \widehat{\nu}_{ab} p_{ab}(j|i;t) 
  \delta_{jj'} \delta_{ji'} (1-\delta_{ij}) \, ,
\end{eqnarray}\reseteqn
where the {\sc Kronecker} {\em symbol} $\delta_{ij}$ is defined by
\begin{displaymath}
 \delta_{ij} := \left\{
\begin{array}{ll}
1 & \mbox{if } i=j \\
0 & \mbox{if } i\ne j \,.
\end{array}\right.
\end{displaymath}                                        
The factors $(1 - \delta_{ij})$ result from the constraint $i \ne j$, whereas
factors of the form $\delta_{ij}$ correspond to conditions of the kind
$i = j$ which follow by comparison of (\ref{transform}a) and (\ref{transform}b)
respectively with (\ref{orgform}). The parameter
\begin{equation}
 \nu_{ab} := N_b \, \widehat{\nu}_{ab}
\label{NUAB}
\end{equation}
represents the {\em contact rate} between an 
individual of subpopulation $a$ with individuals
of subpopulation $b$. $p_{ab}(j|i;t)$ denotes the probability of an 
individual of subpopulation $a$ to change the 
strategy from $i$ to $j$ during an imitative pair interaction
with an individual of subpopulation $b$, i.e.
\begin{displaymath}
 \sum_j p_{ab} (j|i;t) = 1 \, .
\end{displaymath}
For $j \ne i$ we will assume
\begin{equation}
 p_{ab}(j|i;t) := f_{ab} \widehat{R}_a(j|i;t) 
\label{PABK}
\end{equation}
where the parameter $f_{ab}$ is a measure for the {\em frequency}
of imitative pair interactions 
between individuals of subpopulation $a$ when confronted with an individual
of subpopulation $b$. $\widehat{R}_a(j|i;t)$
is a measure for the {\em readiness} 
of individuals belonging to subpopulation $a$ 
to change the strategy from $i$ to $j$ during a pair interaction.

\subsection{`Microscopic' foundation of evolutionary game theory} 
\label{Micfound}
The problem of this section is to specify the frequency $f_{ab}$ and the 
readiness $\widehat{R}_a(j|i;t)\equiv \widehat{R}_a(j|i;\vec{n};t)$ 
in an adequate way. For this we make the following assumptions:
\begin{itemize}
\item By experience each individual knows---at least
approximately---the {\em expected success} 
of the strategy used: 
We will define the expected success of a strategy $i$ for an individual of
subpopulation $a$ in interactions with other individuals by
\begin{equation}
 \widehat{E}_a(i,t) \equiv \widehat{E}_a(i,\vec{n};t) 
 := \sum_b \sum_j r_{ab} E_{ab}(i,j) 
 \frac{n_j^b(t)}{N_b} \, .
\label{formel}
\end{equation}
Here, the parameter
\[
 r_{ab} = \frac{\nu_{ab}}{\displaystyle \sum_c \nu_{ac}} 
\]
represents the {\em relative contact
rate} of an individual of subpopulation $a$
with individuals of subpopulation $b$. 
$n_j^b(t)/N_b$ is the probability that an interaction partner of subpopulation
$b$ uses strategy $j$. 
$E_{ab}(i,j)$ is an exogenously given quantity that
denotes the {\em success} of strategy $i$ for an individual
of subpopulation $a$ during an interaction with an individual of subpopulation
$b$ who uses strategy $j$. 
Since all these quantities can be determined by each
individual, the evaluation of the expected success $\widehat{E}_a(i,t)$ 
is obviously possible. 
\item In interactions with individuals of the 
{\em same} subpopulation an individual
tends to take over the strategy of another individual if the expected success
would increase: 
When an individual who uses strategy $i$ meets another individual
of the same subpopulation who uses strategy $j$, they compare their
expected successes $\widehat{E}_a(i,t)$ and
$\widehat{E}_a(j,t)$ respectively by exchange of their experiences. 
(Remember that individuals of the same subpopulation were 
assumed to cooperate.) The individual with strategy $i$ will {\em imitate}
the other's strategy $j$ with a probability $p_{ab}(j|i;t)$ that is 
growing with the expected increase
\begin{displaymath}
 \Delta_{ji} \widehat{E}_a := \widehat{E}_a(j,t) - \widehat{E}_a(i,t)
\end{displaymath}
of success. If a change of strategy would imply a {\em decrease} of success
($\Delta_{ji} \widehat{E}_a < 0$), 
the individual will not change the strategy $i$.
Therefore, the readiness for replacing the strategy $i$ by $j$ during an
interaction within the same subpopulation can be assumed to be
\begin{equation}
 \widehat{R}_a(j|i;t) := \max \Big( \widehat{E}_a(j,t) 
 - \widehat{E}_a(i,t),0 \Big) 
\label{read}
\end{equation}
where $\max(x,y)$ is the maximum of the two numbers $x$ and $y$.
This describes an individual {\em optimization} or {\em learning process}.
\item In interactions with individuals of {\em other} subpopulations
(who behave in a non-cooperative way) 
normally no imitative processes will take
place: During these interactions the expected success $\widehat{E}_b(j,t)$
of the interaction partner can at best be {\em estimated} by {\em observation}
since he will not tell his experiences. Moreover, due to different criteria
for the grade of success, the expected success of a strategy $j$ will
normally be varying with the subpopulation (i.e. 
$\widehat{E}_a(i,t) \ne \widehat{E}_b(i,t)$ 
for \mbox{$a\ne b$}). As a consequence, an imitation of the strategy of
individuals belonging to {\em another} subpopulation would be very risky since it
would probably be connected with a {\em decrease} of expected success.
Hence the assumption
\begin{equation}
 f_{ab} := \delta_{ab} = \left\{
\begin{array}{ll}
1 & \mbox{if } a=b \\
0 & \mbox{if } a\ne b \,.
\end{array}\right. 
\label{fab}
\end{equation}
will normally be justified.
\par
Relation (\ref{fab}) also results in cases where the strategies of the
respective other subpopulations {\em cannot} be imitated due to
different (disjunct) strategy sets. Then, we need not to assume that
individuals of the same subpopulations cooperate, whereas individuals of
different subpopulations do not. 
\end{itemize}
In Section \ref{dyngame} it will turn out that the game dynamical equations
are the approximate mean value equations of the stochastic behavioral
model defined by (\ref{specform}) to (\ref{fab}).
In this sense, the model of this
section can be regarded as stochastic version of the game dynamical
equations. Moreover, the assumptions made above are a `microscopic' 
foundation of evolutionary game theory since they allow a derivation
of the game dynamical equations on the basis of individual
behavior patterns. 

\subsection{Self-organization of behavioral conventions by competition between
strategies} \label{Selfor}

As an example for the stochastic game dynamical equations 
we will consider a case
with one subpopulation only ($A=1$). 
In this case we can omit the
indices $a$, $b$, and the summation over $b$.
Let us assume the individuals to choose between two
{\em equivalent} strategies $i \in \{1,2\}$, 
i.e. the {\em success matrix}
$\underline{E} \equiv \Big( E(i,j) \Big)$ 
is symmetrical:
\begin{equation}
 \underline{E} :=
\left(
\begin{array}{cc}                                   
B+C & B \\                                                   
B & B+C
\end{array}\right) \, .                                 
\label{pay}
\end{equation}
According to the relation
\[
n_1(t) + n_2(t) = N
\]
(cf. (\ref{sum})), $n_2(t) = N - n_1(t)$ is already
determined by $n_1(t)$. 
For spontaneous strategy changes due to {\em trial and error} 
we will take the simplest form of transition rates:
\begin{equation}
 w(j|i;t) := W \, .
\label{fluct}
\end{equation}
A situation of the above kind is the avoidance behavior of pedestrians (cf.
{\sc Helbing}, 1991, 1992): 
In pedestrian crowds with two opposite directions of motion, 
pedestrians have sometimes to avoid each other in order to exclude a collision.
For an avoidance maneuver to be successful, both pedestrians concerned have to
pass the respective other pedestrian either on the right hand side 
(strategy $i=1$) or on the left hand side (strategy $i=2$). 
Otherwise, both pedestrians have to stop 
(cf. Figure \ref{separation}a).\marginpar{Fig. 1}
Here, both strategies 
are equivalent, but the success of a strategy increases
with the number $n_i$ of individuals who use the {\em same} strategy.
In success matrix (\ref{pay}) we have 
\[
C > 0 \, ,
\] 
then.
\par
Empirically one finds that 
the probability $P_1$ of choosing the right hand side
is usually different from the probability $P_2=1-P_1$ 
of choosing the left hand side. 
Consequently, opposite directions of motion 
normally use separate lanes (cf. Figure~\ref{separation}b).
\par
We will now examine if our behavioral model can explain this {\em 
symmetry breaking} (the fact that $P_1 \ne P_2$). 
Figure~\ref{phasetr} shows some computational results for 
$C=1$ and different values of $W/\nu$.\marginpar{Fig. 2}
If 
\begin{equation}
 \kappa := 1 - \frac{4W}{\nu C} < 0 \, ,
\end{equation}
the configurational distribution is unimodal and symmetrical with respect to
$n_1 = N/2 = n_2$, i.e. both strategies will be chosen by about one
half of the individuals. 
At the {\em critical point} $\kappa = 0$
there appears a {\em phase transition (bifurcation)}. This is indicated by the
broadness of the probability distribution 
$P(\vec{n},t) \equiv P(n_1,n_2;t) = P(n_1,N-n_1;t)$ 
which comes from socalled {\em critical
fluctuations} (cf. {\sc Haken}, 1983). The term `critical fluctuations'
denotes the fact that the fluctuations become particularly large at a critical
point since the system behavior is unstable, then. Whereas the individuals
behave more or less independently {\em before} the phase transition
($\kappa < 0$), around the critical point the individuals begin
to act correlated due to their (imitative) interactions. However, the
spontaneous strategy changes (represented by $W$) still prevent the formation 
of a behavioral preference. Above the critical point (i.e. for $\kappa > 0$)
the correlation of individual behaviors is strong enough for the 
{\em self-organization (emergence) of a behavioral convention:}
The configurational distribution becomes multimodal 
in the course of time with maxima at $n_1 \ne N/2$ so that
one of the two equivalent strategies will very probably be chosen by
a {\em majority} of individuals. In this connection one also speaks of
{\em symmetry breaking} ({\sc Haken}, 1979, 1983).
\par
Behavioral conventions often obtain a law-like character
after some time. Which one of two equivalent strategies will win 
the majority is completely
random. It is possible, that conventions differ from one 
region to another. This is, for example, the case for the prescribed
driving direction of cars.
\par
The model of this section can also be applied to 
the competition between the two video systems
VHS and BETA MAX which were equivalent with respect to technology and price 
at the beginning ({\sc Hauk}, 1994). 
In the course of time VHS won this rivalry since (for
reasons of {\em compatibility}
concerning copying, selling or hiring of video tapes) it was advantageous
for new purchasers to decide for that video
system which gained a small majority at
some moment. Other examples for the emergence of a behavioral convention are
the revolution direction of clock hands, the direction of writing, etc.
A generalization of the above model to the case of more than two alternative
strategies is easily possible. Of course, the model can also be adapted
to situations where one behavioral alternative is superior to the others.
However, the formation of a behavioral convention is trivial, then.
\par
Finally, some related models should be mentioned which were proposed
during the recent years for the description of symmetry breaking 
phenomena in economics: {\sc Orl\'{e}an} (1992, 1993) and
{\sc Orl\'{e}an} and {\sc Robin} (1992) presented a phase transition model
using polynomial transition rates which base on a {\sc Bayes}ian rationale.
{\sc Durlauf} (1989, 1991) used {\sc Markov}ian fields to explain the
non-ergodic (i.e. path-dependent) behavior of some economic systems.
{\sc F\"ollmer} (1974) applied the {\sc Ising} model paradigm (1925)
to model an economy of many interacting agents and discussed under which
conditions a symmetry breakdown occurs. A similar model for polarization
effects in opinion formation was already suggested by {\sc Weidlich} (1972).
Last but not least {\sc Topol} (1991) presented a {\sc Fokker-Planck} equation
model for the explanation of bubbles in stock markets by mimetic contagion
(i.e. some kind of imitative interactions) between agents.

\section{Most probable and expected strategy distribution} \label{Mostexp}

Because of the huge number of possible socioconfigurations $\vec{n}$,
in more complex cases than in Section \ref{Selfor}
the master equation for the determination of the configurational distribution
$P(\vec{n},t)$ is usually difficult to solve (even with a computer).
However,
\begin{itemize}
\item in cases of the description of single or rare social 
processes the {\em most
probable} strategy distribution
\begin{equation}
 P_i^a(t) := \frac{\widehat{n}_i^a(t)}{N_a}
\end{equation}
is the quantity of interest whereas
\item in cases of frequently occuring social processes the interesting
quantity is the {\em expected} strategy distribution
\begin{equation}
 P_i^a(t) := \frac{\langle n_i^a \rangle_t}{N_a} \, .
\label{MeaN}
\end{equation}
\end{itemize}
$P_i^a(t)$ is the {\em proportion} of individuals within subpopulation $a$
using strategy $i$ so that 
\[
 P_i^a(t) \ge 0 \qquad \mbox{and} \qquad \sum_i P_i^a(t) = 1 \, .
\]
Equations for the {\em most probable occupation numbers}
$\widehat{n}_i^a(t)$ can be deduced from a 
{\sc Langevin} equation (1908)
for the temporal development of the socioconfiguration $\vec{n}(t)$. 
For the {\em mean values} $\langle n_i^a
\rangle_t$ {\em of the occupation numbers}
$n_i^a$ normally only {\em approximate} closed equations can be derived.
A measure for the reliability of $\widehat{n}_i^a(t)$
and $\langle n_i^a \rangle_t$ with respect to the possible temporal
developments of $n_i^a(t)$ are the variances $\sigma_{ii}^{aa}(t)$ of 
$n_i^a(t)$. If the standard deviation $\sqrt{\sigma_{ii}^{aa}(t)}$ becomes
comparable to $0.12 \widehat{n}_i^a(t)$ or $0.12 \langle n_i^a \rangle_t$,
the values of $\widehat{n}_i^a(t)$ and $\langle n_i^a \rangle_t$
respectively are not representable for $n_i^a(t)$ any more
(cf. Section \ref{CompRes}). In the case of
$P(\vec{n},t)$ being normally distributed this would imply a probability
of 34\% (5\%) that the value of $n_i^a(t)$ deviated more than 12\% (24\%)
from $\widehat{n}_i^a(t)$ and $\langle n_i^a \rangle_t$ respectively. Moreover,
if the variances $\sigma_{ii}^{aa}(t)$ become large, this may indicate a phase
transition, i.e. a non-ergodic (path-dependent) temporal evolution of the
system (see Figure \ref{phasetr}).

\subsection{Mean value and covariance equations}

The {\em mean value}
of a function $f(\vec{n},t)$ is defined by
\begin{displaymath}
 \langle f(\vec{n},t) \rangle_t \equiv
 \langle f(\vec{n},t) \rangle := \sum_{\weg{n}} f(\vec{n},t) P(\vec{n},t) \, .
\end{displaymath}
From master equation (\ref{MAster}) can be derived
that the mean values of the occupation 
numbers $f(\vec{n},t) = n_i^a$
are determined by the equations
\begin{equation}
 \frac{d\langle n_i^a\rangle}{dt} = \langle m_i^a(\vec{n},t) \rangle
\label{mean}
\end{equation}
with the {\em drift coefficients}
\begin{eqnarray}
 m_i^a(\vec{n},t) &:=& \sum_{\weg{n}'} (n'{}_i^a - n_i^a)w(\vec{n}'|\vec{n};t)
\nonumber \\
&=& \sum_{i'} \Big[ \overline{w}^a(i|i';t)n_{i'}^a 
 -  \overline{w}^a(i'|i;t) n_i^a \Big]
\label{mean1}
\end{eqnarray}
and the {\em effective transition rates}
\begin{equation}
 \overline{w}^a(i'|i;t) := \widehat{w}_a(i'|i;t) 
+ \sum_b \sum_{j'}\sum_{j}
\widehat{w}_{ab}(i',j'|i,j;t)n_{j}^b 
\label{mean2}
\end{equation}
(cf. {\sc Helbing}, 1992, 1992a). Obviously, the contributions 
$\widehat{w}_{ab}(i',j'|i,j;t)n_{j}^b$ 
due to pair interactions are proportional
to the number $n_{j}^b$ of possible interaction partners.

\subsubsection{Approximate mean value equations}

Equations (\ref{mean}) are no closed equations, since they depend on the
mean values $\langle n_i^a n_j^b \rangle$, which are not determined
by (\ref{mean}). We have, therefore, to find a suitable approximation.
Using a {\em first order} {\sc Taylor} approximation we obtain
the {\em approximate mean value equations}
\begin{eqnarray}
 \frac{\partial \langle n_i^a \rangle}{\partial t}
&\approx& \Bigg\langle m_i^a(\langle \vec{n} \rangle,t) 
+ \sum_{b,j} (n_j^b - \langle n_j^b \rangle )
\frac{\partial m_i^a(\langle \vec{n} \rangle,t)}{\partial \langle n_j^b\rangle}
\Bigg\rangle \nonumber \\
&=& m_i^a(\langle\vec{n}\rangle,t) \, .
\label{Mean}
\end{eqnarray}
These are applicable if the configurational 
distribution $P(\vec{n},t)$ has only
small covariances 
\begin{eqnarray}
 \sigma_{ij}^{ab} &:=& \Big\langle (n_i^a - \langle n_i^a \rangle)
 (n_j^b - \langle n_j^b \rangle) \Big\rangle \nonumber \\
&=& \langle n_i^a n_j^b \rangle - \langle n_i^a \rangle \langle n_j^b \rangle
 \approx 0 \, . \label{Covar}
\end{eqnarray}
Condition (\ref{Covar}) corresponds to the limit of 
{\em statistical independence} $\langle n_i^a n_j^b \rangle 
= \langle n_i^a \rangle \langle n_j^b \rangle$ of the occupation numbers
(and, therefore, of the individual behaviors).

\subsubsection{{\sc Boltzmann}-like equations}

Inserting (\ref{MeaN}), (\ref{mean1}), and (\ref{mean2}) 
into (\ref{Mean}) the resulting approximate 
equations for the expected strategy distribution $P_i^a(t)$ are
\begin{equation}
 \frac{d}{dt} P_i^a(t) = \sum_{i'} \Big[ w^a(i|i';t)P_{i'}^a(t)
 - w^a(i'|i;t) P_i^a(t) \Big]
\label{boltz1}
\end{equation}
with the {\em mean transition rates}
\begin{equation}
 w^a(i'|i;t) = \widehat{w}_a(i'|i;t) + \sum_b \sum_{j'} \sum_j
 N_b \, \widehat{w}_{ab}(i',j'|i,j;t) P_j^b(t) \, .
\label{boltz2}
\end{equation}
Equations (\ref{boltz1}), (\ref{boltz2}) 
are called {\sc Boltzmann}-{\em like equations}
({\sc Boltzmann}, 1964; {\sc Helbing}, 1992, 1992a) since
the mean transition rates (\ref{boltz2}) depend on the strategy distributions
$P_j^b(t)$ due to pair interactions.
Assuming (\ref{specform}), (\ref{NUAB}), and (\ref{PABK}) 
we obtain the formula
\begin{equation}
 w^a(i|i';t) = \widehat{w}_a(i|i';t) + 
 R_a(i|i';t) \sum_b \nu_{ab} f_{ab} P_i^b(t) 
\label{boltz3}
\end{equation}
with $R_a(i|i';t) := \widehat{R}_a(i|i';\langle \vec{n} \rangle ; t)$ 
for the mean transition rates.
(\ref{boltz1}) and (\ref{boltz3}) are a special case of more general equations 
introduced by {\sc Helbing} (1992, 1992b, 1994)
for the temporal development of the expected strategy distribution
in a social system consisting of a huge number $N \gg 1$ of individuals.

\subsubsection{Approximate covariance equations}

In many cases, the configuration $\vec{n}_0$ at an initial time $t_0$ is 
known by empirical evaluation, i.e. the initial distribution is
\begin{displaymath}
 P(\vec{n},t_0) = \delta_{\weg{n}\weg{n}_0} \, .
\end{displaymath}
As a consequence, the covariances $\sigma_{ij}^{ab}$ vanish at time 
$t_0$ and remain small during a certain time interval. For the temporal
development of $\sigma_{ij}^{ab}$, the equations 
\begin{equation}
 \frac{d\sigma_{ij}^{ab}}{dt} = \Big\langle m_{ij}^{ab}(\vec{n},t)
\Big\rangle 
+ \Big\langle (n_i^a - \langle n_i^a \rangle )m_j^b(\vec{n},t)
\Big\rangle 
+ \Big\langle (n_j^b - \langle n_j^b \rangle) m_i^a(\vec{n},t)
\Big\rangle
\label{cov}
\end{equation}
can be derived from master equation (\ref{MAster}) (cf. {\sc Helbing},
1992, 1992a). Here,
\begin{eqnarray}
 m_{ij}^{ab}(\vec{n},t) 
&:=& \sum_{\weg{n}'} (n'{}_i^a - n_i^a) 
 (n'{}_j^b - n_j^b) w(\vec{n}'|\vec{n};t) \nonumber \\
&=& \delta_{ab} \bigg( \delta_{ii'}
\sum_{j}\Big[ n_{j}^a \overline{w}^a(i|j;t) 
+ n_{i}^a \overline{w}^a(j|i;t) \Big] \nonumber \\
& & \qquad - \; \Big[n_{i'}^a \overline{w}^a(i|i';t) 
+ n_{i}^a \overline{w}^a(i'|i;t) \Big]\bigg) \nonumber \\
&+& \sum_{j'}\sum_{j} \Big[                        
n_{j}^a n_{j'}^b \widehat{w}_{ab}(i,i'|j,j';t) 
+ n_{i}^a n_{i'}^b \widehat{w}_{ab}
(j,j'|i,i';t) \Big] \vphantom{\sum_j} \nonumber \\
&-& \sum_{j'}\sum_{j} \Big[
n_{i}^a n_{j'}^b  \widehat{w}_{ab}(j,i'|i,j';t) 
+  n_{j}^an_{i'}^b \widehat{w}_{ab}(i,j'|j,i';t) 
\Big]  \qquad
\label{cov1}
\end{eqnarray}
are {\em diffusion coefficients}. 
Equations (\ref{cov}) are again no closed equations. However, a first order
{\sc Taylor} approximation of the drift
and diffusion coefficients $m_{..}^{..}(\vec{n},t)$
leads to the equations
\begin{equation}
 \frac{\partial \sigma_{ij}^{ab}}{\partial t} 
\approx m_{ij}^{ab}(\langle n \rangle,t)              
+ \sum_{c,k} \Bigg( \sigma_{ik}^{ac} \frac{\partial m_j^b(\langle \vec{n}
\rangle,t)}{\partial \langle n_k^c \rangle} 
+ \sigma_{jk}^{bc} \frac{\partial m_i^a(\langle \vec{n}\rangle,t)}
{\partial \langle n_k^c \rangle} \Bigg) \qquad
\label{Cov}
\end{equation}
(cf. {\sc Helbing}, 1992, 1992a) which are solvable 
together with (\ref{Mean}).
The {\em approximate covariance equations} 
(\ref{Cov}) allow the determination of
the time interval during which the approximate mean value equations
(\ref{Mean}) are valid (cf. Section \ref{CompRes} and Figures \ref{valid1}, 
\ref{valid2}). They are also useful for the calculation of the
reliability (or representativity) of descriptions made by 
equations (\ref{Mean}). Moreover, they are necessary for {\em corrections} of 
approximate mean value equations (\ref{Mean}).
    
\subsubsection{Corrected mean value and covariance equations} \label{Corr}

Equations (\ref{Mean}) and (\ref{Cov}) are only valid for the case 
\begin{equation}
 \Big| \sigma_{ij}^{ab} \Big| \ll \langle n_i^a \rangle \langle n_j^b \rangle 
\label{sm}
\end{equation}
where the absolute values of the cova\-ri\-an\-ces $\sigma_{ij}^{ab}$
are small, i.e. where the 
configurational distribution $P(\vec{n},t)$ is
sharply peaked. For increasing covariances a better approximation
of (\ref{mean}) and (\ref{cov}) should be taken. A {\em second order} 
{\sc Taylor} approximation of (\ref{mean}) and (\ref{cov}) respectively
results in the {\em corrected mean value equations}
\begin{equation}
 \frac{\partial \langle n_i^a \rangle}{\partial t}
\approx m_i^a(\langle \vec{n} \rangle,t) 
+ \frac{1}{2}
\sum_{b,j}\sum_{c,k} \sigma_{jk}^{bc} \frac{\partial^2 m_i^a(\langle
\vec{n} \rangle,t)}{\partial \langle n_j^b\rangle\partial \langle
n_k^c \rangle} \qquad 
\label{corrmean}
\end{equation}
and the {\em corrected covariance equations}
\begin{eqnarray}
 \frac{d\sigma_{ij}^{ab}}{dt} &\approx& m_{ij}^{ab}(\langle n \rangle,t)
+ \frac{1}{2} \sum_{c,k} \sum_{d,l} \sigma_{kl}^{cd} \frac{
\partial^2 m_{ij}^{ab}(\langle\vec{n}\rangle,t)}{\partial \langle n_k^c
\rangle \partial \langle n_l^d \rangle} \qquad \nonumber \\
&+& \sum_{c,k} \Bigg( \sigma_{ik}^{ac} \frac{\partial m_j^b(\langle \vec{n}
\rangle,t)}{\partial \langle n_k^c \rangle} 
+ \sigma_{jk}^{bc} 
\frac{\partial m_i^a(\langle \vec{n}\rangle,t)}
{\partial \langle n_k^c \rangle} \Bigg) 
\label{corrcov}
\end{eqnarray}
({\sc Helbing}, 1992, 1992a).
Note that the corrected mean value equations explicitly depend on the
covariances $\sigma_{ij}^{ab}$, i.e.
on the {\em fluctuations} due to the stochasticity of
the processes described. They cannot be solved without
solving the covariance equations. A comparison of (\ref{corrmean}) with
(\ref{Mean}) shows that the approximate mean value equations only agree with the
corrected ones in the limit $\sigma_{ij}^{ab}$ of negligible corvariances
(cf. also (\ref{Covar})).
However, the calculation of the
covariances is {\em always} recommendable since they are a measure for the
reliability (or representativity) of the mean value equations.
If the covariances become large in the
sense of equation (\ref{crIT}) this may indicate a phase transition.

\subsubsection{Computational results} \label{CompRes}

A comparison of exact, approximate and corrected mean value and variance
equations is 
given in Figures \ref{below2} to \ref{valid}a.
These show computational results
corresponding to the example of Section \ref{Selfor} (cf. Figure
\ref{phasetr}).
{\em Exact} mean values $\langle n_1 \rangle$
and variances $\sigma_{11}$ 
are represented by solid lines whereas approximate
results according to (\ref{Mean}), (\ref{Cov})       
are represented by dotted
lines and corrected results according to (\ref{corrmean}), (\ref{corrcov})
by broken lines.\marginpar{Fig. 3--4}
\par
For $\kappa \ge 0$ the approximate mean value equations (\ref{Mean})
become useless since the variances are growing due to
the {\em phase transition}. 
As expected, the corrected mean value equations yield better
results than the approximate mean value equations and they are valid
for a longer time interval. 
\par
A criterium for the validity of the approximate equations 
(\ref{Mean}), (\ref{Cov}) and the
corrected equations (\ref{corrmean}), (\ref{corrcov}) respectively
are the {\em relative central moments}
\[
C_m(t) \equiv C_{i_1}^{a_1}{}_{\dots}^{\dots}{}_{i_m}^{a_m}(t) 
:= \frac{\Big\langle (n_{i_1}^{a_1} - \langle
n_{i_1}^{a_1} \rangle ) \cdot \dots \cdot (n_{i_m}^{a_m} - \langle
n_{i_m}^{a_m} \rangle ) \Big\rangle }{\langle n_{i_1}^{a_1} \rangle
\cdot  \dots  \cdot \langle n_{i_m}^{a_m} \rangle} \, .
\]
Whereas the approximate 
equations (\ref{Mean}), (\ref{Cov}) already fail, if
\begin{equation}
 |C_m(t) | \le 0.04
\label{Bed}
\end{equation}
is violated for $m=2$
(compare to (\ref{sm}), (\ref{Covar})), the corrected equations
(\ref{corrmean}), (\ref{corrcov}) presuppose condition (\ref{Bed})
only for $3 \le m \le l$ with a certain, well-defined value $l$ 
(cf. {\sc Helbing}, 1992, 1992a for details). However, even
the corrected equations (\ref{corrmean}), (\ref{corrcov}) become useless
if the probability distribution $P(\vec{n},t)$ becomes multimodal, i.e.
if a phase transition occurs. This is the case if 
\begin{equation}
 |C_2(t) | = \left| \frac{\sigma_{ij}^{ab}(t)}{\langle n_i^a \rangle
 \langle n_j^b \rangle} \right| \le 0.12
\label{crIT}
\end{equation}
is violated (cf. Figure \ref{valid}).\marginpar{Fig. 5a,b}

\subsection{Equations for the most probable strategy distribution}
\label{Most}          
After the transformation of master equation (\ref{master}) 
into a {\sc Fokker-Planck}
equation by a second order {\sc Taylor} approximation, it can be
reformulated in terms of a {\sc Langevin} {\em equation} (1908) 
(cf. {\sc Weidlich} and {\sc Haag}, 1983; {\sc Helbing}, 1992). 
The latter reads
\begin{equation}
 \frac{d}{dt} n_i^a(t) \stackrel{N_a \gg 1}{=} m_i^a(\vec{n},t) +
 \mbox{\em fluctuations} 
\label{Lang}
\end{equation}
and describes the temporal
development of the 
socioconfiguration $\vec{n}(t)$ in dependence of 
process immanent fluctuations
(that are determined by the diffusion coefficients $m_{ij}^{ab}$). As a 
consequence, 
\begin{equation}
\frac{d}{dt} \widehat{n}_i^a(t) \stackrel{N_a\gg 1}{=} 
m_i^a(\widehat{\vec{n}},t)
\label{mostprob}
\end{equation}
are the equations governing the temporal development of
the {\em most probable} occupation numbers
$\widehat{n}_i^a(t)$. Equations (\ref{mostprob}) look exactly like
approximate mean value equations (\ref{Mean}). Therefore,
if $N_a \gg 1$, the {\em approximate}
mean value equations have an interpretation even for
large variances since they also describe the most probable strategy
distribution. 

\section{The game dynamical equations} \label{dyngame}

\subsection{`Macroscopic' derivation} \label{Macr}

Before we will connect the stochastic behavioral model to the
game dynamical equations, we will discuss their derivation from a
collective level of behavior.
Let $E_a(i,t) := \widehat{E}_a(i,\langle \vec{n} \rangle;t)$ 
be the {\em expected success} of strategy $i$ for
an individual of subpopulation $a$ and
\begin{equation}
 \overline{E_a}(t) := \sum_i E_a(i,t)P_i^a(t)
\label{EA}
\end{equation}
the {\em mean expected success}. If the {\em relative} increase
\begin{displaymath}
 \frac{dP_i^a/dt}{P_i^a(t)}
\end{displaymath}
of the proportion $P_i^a(t)$ is assumed to be proportional to the difference
$[E_a(i,t) - \overline{E_a}(t) ]$ between 
the expected and the mean expected success, one obtains
the {\em game dynamical equations}
\begin{equation}
 \frac{d}{dt} P_i^a(t) = \nu_a P_i^a(t) \Big[ E_a(i,t) - 
 \overline{E_a}(t) \Big] \, .
\label{sel}
\end{equation}
According to these equations
the proportions of strategies with an expected success 
that exceeds the {\em average} $\overline{E_a}(t)$ are growing, 
whereas the proportions
of the remaining strategies are falling.
For the expected success $E_a(i,t)$ one often takes the form
\begin{equation}
 E_a(i,t) := \sum_b \sum_j A_{ab}(i,j) P_j^b(t) 
\label{payoff}
\end{equation}
where the quantities $A_{ab}(i,j)$ have the meaning of 
{\em payoffs} which are exogeneously determined. 
Consequently, the matrices
\[
\underline{A}_{ab} := \Big( A_{ab}(i,j) \Big)
\]
are called {\em payoff matrices}.
Inserting (\ref{EA}) and (\ref{payoff}) 
into (\ref{sel}), one obtains the explicit form
\begin{equation}
 \frac{d}{dt} P_i^a(t) 
= \nu_a P_i^a(t) \Bigg[ \sum_{b,j} A_{ab}(i,j)
P_j^b(t) 
- \sum_{i'} \sum_{b,j} P_{i'}^a(t)A_{ab}(i',j)P_j^b(t) \Bigg] 
\label{game}
\end{equation}                                                     
of the game dynamical equations. 
Equations of this kind are
very useful for the investigation and understanding of the competition or
cooperation of individuals (cf. e.g. 
{\sc Hofbauer} and {\sc Sigmund}, 1988;
{\sc Schuster} et.al., 1981). 
Due to their nonlinearity
they may have a complex dynamical solution, 
e.g. an {\em oscillatory} one
({\sc Hofbauer} et.al., 1980; {\sc Hofbauer} and {\sc Sigmund}, 1988)
or even a {\em chaotic} one ({\sc Schnabl} et.al., 1991). 
\par
A slightly generalized form of 
(\ref{sel}),\alpheqn{Game}
\begin{eqnarray}
\frac{d}{dt} P_i^a(t) 
&=& \sum_{i'} \Big[ \widehat{w}_a(i|i';t)P_{i'}^a(t) 
- \widehat{w}_a(i'|i;t)P_i^a(t) \Big] \label{mutation} \\
&+& \nu_a P_i^a(t) \Big[ E_a(i,t) - \overline{E_a}(t) \Big] \, , \qquad 
\label{selection}
\end{eqnarray}\reseteqn
is also known as {\em selection mutation equation} 
({\sc Hofbauer} and {\sc Sigmund}, 1988):
(\ref{selection}) can be understood as effect of a {\em selection}
(if $E_a(i,t)$ is interpreted as {\em fitness} of strategy~$i$), and
(\ref{mutation}) can be understood as effect of {\em mutations}. Equation
(\ref{Game}) is a powerful tool in 
evolutionary biology (cf. {\sc Eigen}, 1971; {\sc Fisher}, 1930;
{\sc Eigen} and {\sc Schuster}, 1979; {\sc Hofbauer} and {\sc Sigmund}, 
1988; {\sc Feistel} and {\sc Ebeling}, 1989).
In game theory, the mutation term could be used for the description
of {\em trial and error behavior} or accidental variations of the strategy.

\subsection{Derivation from the stochastic behavioral model} \label{Deriv}

In this section we will look for a connection between the
stochastic behavioral model of Section \ref{Stogame}
and the game dynamical equations. For this purpose we
compare the approximate mean value equations
of this stochastic behavioral model, i.e. the
{\sc Boltzmann}-like equations (\ref{boltz1}), (\ref{boltz3})
with the game dynamical
equations (\ref{Game}). Both equations will be identical only if
\[
 \nu_{ab}f_{ab} = \nu_a \delta_{ab} \, .
\]
This condition corresponds to (\ref{fab}) if
\[
 \nu_a = \nu_{aa} \, .
\]
Inserting assumptions (\ref{formel}) to (\ref{fab}) into the 
{\sc Boltzmann}-like equations (\ref{boltz1}), (\ref{boltz3})
the game dynamical equations (\ref{Game}) result. We have
only to introduce the identifications
\[
 A_{ab}(i,j) := r_{ab} E_{ab}(i,j) \, ,
\]
\[
 E_a(i,t) := \widehat{E}_a(i,\langle \vec{n} \rangle ; t) 
= \sum_b \sum_j r_{ab} E_{ab}(i,j) P_j^b(t) \, ,
\]
and to apply the relation
\[
 \max \Big( E_a(i,t)-E_a(j,t),0 \Big) 
- \max \Big( E_a(j,t) - E_a(i,t),0 \Big)                    
= E_a(i,t) - E_a(j,t) \, .
\]
The game dynamical equations (including their properties and generalizations)
are more explicitly discussed 
elsewhere ({\sc Helbing}, 1992). An interesting application to a case with
two subpopulations can be found in the book of {\sc Hofbauer} and
{\sc Sigmund} (1988: pp. 137--146). In the following,
we will again examine the example of Section \ref{Selfor} where
we have one subpopulation and two equivalent strategies.
The game dynamical equations (\ref{Game}) corresponding
to (\ref{pay}) and (\ref{fluct}) have, then, the explicit form
\begin{equation}
 \frac{d}{dt} P_i(t) = -2\left( P_i(t) - \frac{1}{2} \right) 
\Big[ W + \nu C P_i(t) \Big( P_i(t) - 1 \Big) \Big] \, . 
\label{concr}
\end{equation}
According to (\ref{concr}), $P_i = 1/2$
is a stationary solution. This solution is stable for
\[
 \kappa = 1 - \frac{4W}{\nu C} 
 < 0 \, ,
\]
i.e. if spontaneous strategy changes are dominating and, therefore, 
prevent a self-organ\-iza\-tion process.
\par
At the {\em critical point} $\kappa = 0$ 
{\em symmetry breaking} appears:                  
For $\kappa > 0$ the stationary solution $P_i = 1/2$ is unstable and
the game dynamical equations (\ref{concr}) can be rewritten in the form
\begin{equation}
 \frac{d}{dt} P_i(t) = -2 \left( P_i(t) - \frac{1}{2} \right) 
\left( P_i(t) - \frac{1 + \sqrt{\kappa}}{2} \right) 
\left( P_i(t) - \frac{1-\sqrt{\kappa}}{2} \right) \, .
\end{equation}
This means, for $\kappa > 0$ we have two additional stationary solutions 
$P_i = (1+\sqrt{\kappa})/2$ and 
$P_i = (1-\sqrt{\kappa})/2$ which
are stable. Depending on initial fluctuations,
one strategy
will win a majority of $100\cdot\sqrt{\kappa}$ 
percent. This majority is the greater
the smaller the rate $W$ of spontaneous strategy changes is.

\section{Modified game dynamical equations}

At first glance the crease of $P(n_1,N-n_1;t)$ at 
$n_1 = N/2 = n_2$ in the illustrations of Figure~\ref{phasetr} 
appears somewhat surprising. A mathematical analysis shows that this
is a consequence of the crease of the function
$\widehat{R}_a(j|i;t) = \max 
(\widehat{E}_a(j,t) - \widehat{E}_a(i,t),0)$. It
can be avoided by using the modified approach
\begin{equation}
 \widehat{R}_a(j,i;t) 
:= \frac{1}{2} \exp [ \widehat{E}_a(j,t)
 - \widehat{E}_a(i,t) ] \, .
\label{newan}
\end{equation}
(\ref{newan}) also leads to a phase transition
for $\kappa = 0$ (cf. Figure \ref{crit}) and very similar
results for the approximate mean value equations since the
game dynamical equations result as {\sc Taylor} approximation of 
those.\marginpar{Fig. 6}
According to (\ref{newan}), imitative strategy changes from $i$ to $j$
will again occur the more frequent the
greater the expected increase 
$\Delta_{ji} \widehat{E}_a = \widehat{E}_a(j,t) - \widehat{E}_a(i,t)$
of success is. 
\par
Approach (\ref{newan}) originally stems from physics where
the exponential function for the transition probability is due to the need to
obtain the {\sc Boltzmann} distribution (1964) as stationary distribution.
Its application to behavioral changes was suggested by {\sc Weidlich}
(1971, 1972) in connection with a {\sc Ising}-like (1925) opinion formation
model. Meanwhile, related models were also proposed for economic systems
({\sc Haag} et. al., 1993; {\sc Weidlich} and {\sc Braun}, 1992; 
{\sc Durlauf}, 1989, 1991). In contrast to this, {\sc Orl\'{e}an}
(1992, 1993) and {\sc Orl\'{e}an} and {\sc Robin} (1992) prefer a transition
probability which has the form of a polynomial of degree two and bases on a
{\sc Bayes}ian rationale.
\par
The advantage of (\ref{newan}) is that it
guarantees the non-negativity of $\widehat{R}_a(j|i;t)$. Moreover, the
exponential approach factorizes into a {\em pull-term}
$\exp[\widehat{E}_a(j,t)]$ and a {\em push-term} $\exp[-\widehat{E}_a(i,t)]$.
Relevant for strategy changes is not the {\em absolute} success 
$\widehat{E}_a(j,t)$ of an available strategy $j$ but the {\em relative} 
success $[\widehat{E}_a(j,t) - \widehat{E}_a(i,t)]$ with respect to the pursued
strategy $i$. 
\par
Furthermore, approach (\ref{newan}) can be related to a decision theoretical
model for {\em choice under risk}. For this let us assume that the {\em utility}
of a strategy change from $i$ to $j$ is given by a known part
\[
 U_a(j|i;t) := \Big[ \widehat{E}_a(j,t) - \widehat{E}_a(i,t) \Big]
\]
and an unknown part $\epsilon_j$ (i.e. an error term) which comes from the
{\em uncertainty} about the exact value of 
$[\widehat{E}_a(j,t) - \widehat{E}_a(i,t)]$ (since $\widehat{E}_a(i,t)$
like $n_j^b$ is subject to fluctuations). If the individual choice
behavior is the result of a maximization process (i.e. if an individual 
chooses the alternative $j$ for which 
$U_a(j|i;t) + \epsilon_j > U_a(i'|i;t) + \epsilon_{i'}$ holds in comparison
with all other available alternatives $i'$) and if the error terms are
identically and independently {\sc Weibull} distributed, the choice
probabilities $p_a(j|i;t)$ are given by the well-known {\em
multinomial logit model} ({\sc Domencich} and
{\sc McFadden}, 1975). It reads
\[
 p_a(j|i;t) = \frac{\exp [\widehat{E}_a(j,t) - \widehat{E}_a(i,t)]}
{\displaystyle \sum_{i'} \exp [\widehat{E}_a(i',t) - \widehat{E}_a(i,t)] }
\, .
\]
(For a more detailed discussion cf. {\sc Helbing}, 1992.)
\par
Approach (\ref{newan}) can also be derived by {\em entropy maximization}
({\sc Helbing}, 1992)
or from the {\em law of relative effect}
in combination with the {\sc Fechner}{\em ian law of psychophysics} 
({\sc Luce}, 1959; {\sc Helbing}, 1992).

\section{Summary and Outlook}

A quite general model for changes of behavioral strategies
has been developed which takes
into account spontaneous changes and changes due to pair interactions.
Three kinds of pair interactions can be distinguished:
imitative, avoidance and compromising processes.
The game dynamical equations result for a special case of imitative processes.
They can be interpreted as equations for the most probable strategy
distribution or as approximate mean value equations 
of a stochastic version of evolutionary
game theory. 
In order to calculate correction terms for the game dynamical equations
as well as to determine the reliability or the time period of validity of game
dynamical descriptions, one has to evaluate the corresponding covariance
equations. Therefore, covariance equations have been derived for a 
very general class of master equations.
\par
The model can be extended in a way that takes into account the expectations
about the future temporal evolution of the expected successes
$E_a(i,t)$ (the {\em `shadow of the future'}). For this purpose,
in (\ref{Game}) $E_a(i,t)$ must be replaced by a quantity $E_a^*(i,t)$ which
represents the expectations about the future success of strategy $i$ on the
basis of its success $E_a(i,t')$ at past times $t' \le t$.  Different
ways of mathematically specifying the future expectations $E_a^*(i,t)$ 
were discussed by {\sc Topol} (1991), {\sc Glance} and {\sc Huberman}
(1992) as well as {\sc Helbing} (1992).

\section*{Acknowledgements}
The author wants to thank Prof. Dr. W. Weid\-lich, 
PD Dr. G. Haag, and Dr. R. Sch\"u\ss{}ler for inspiring discussions.

II. Institute of Theoretical Physics\\
Pfaffenwaldring 57/III\\
70550 Stuttgart\\
Germany
\pagebreak
\begin{figure}[htbp]
\unitlength1cm
\begin{picture}(15,6)(-9,-0.8)
\thicklines
\put(-0.6,-0.2){\line(0,1){6.4}}
\put(3.6,-0.2){\line(0,1){6.4}}
\thinlines
\put(0.5,-0.5){$\Downarrow$}
\put(2.5,6.2){$\Uparrow$}
\put(-1.5,6){(b)}  
\put(0,5.6){\circle{0.4}}
\put(0.4,5.2){\circle{0.4}}
\put(-0.3,5.2){\circle{0.4}}
\put(0.3,4.7){\circle{0.4}}
\put(-0.1,4.3){\circle{0.4}}
\put(0.2,3.9){\circle{0.4}}
\put(0.2,3.9){\vector(1,-2){0.25}}
\put(0,3.4){\circle*{0.4}}
\put(0,3.4){\vector(-1,2){0.25}}
\put(-0.3,3){\circle{0.4}}
\put(0.1,2.5){\circle{0.4}}
\put(-0.2,2){\circle{0.4}}
\put(0.3,1.6){\circle{0.4}}
\put(-0.3,1.3){\circle{0.4}}
\put(0,0.8){\circle{0.4}}
\put(-0.3,0.4){\circle{0.4}}
\put(0.4,0.2){\circle{0.4}}
\put(1.1,0.3){\circle{0.4}}
\put(0.8,0.8){\circle{0.4}}
\put(1.1,1.3){\circle*{0.4}}
\put(1.1,1.3){\vector(1,2){0.25}}
\put(0.9,1.8){\circle{0.4}}
\put(0.9,1.8){\vector(-1,-2){0.25}}
\put(0.8,2.4){\circle{0.4}}
\put(1.5,2.8){\circle*{0.4}}
\put(1.5,2.8){\vector(1,2){0.25}}
\put(0.8,3.2){\circle{0.4}}
\put(1,3.7){\circle{0.4}}
\put(0.7,4.1){\circle{0.4}}
\put(1.3,4.2){\circle{0.4}}
\put(1,4.6){\circle{0.4}}
\put(1.2,5){\circle{0.4}}
\put(0.8,5.4){\circle{0.4}}
\put(1.1,5.9){\circle{0.4}}
\put(1.7,5.6){\circle{0.4}}
\put(1.7,5.6){\vector(-1,-2){0.25}}
\put(2.4,5.5){\circle*{0.4}}
\put(1.8,5.1){\circle*{0.4}}
\put(1.8,5.1){\vector(1,2){0.25}}
\put(2.3,4.8){\circle*{0.4}}
\put(1.9,4.3){\circle*{0.4}}
\put(2.2,3.8){\circle*{0.4}}
\put(1.4,3.4){\circle{0.4}}
\put(1.4,3.4){\vector(-1,-2){0.25}}
\put(2.6,2.3){\circle*{0.4}}
\put(2,2.6){\circle*{0.4}}
\put(1.7,2.2){\circle*{0.4}}
\put(2.3,1.7){\circle*{0.4}}
\put(1.9,1.3){\circle*{0.4}}
\put(2.2,0.9){\circle*{0.4}}
\put(1.5,0.8){\circle*{0.4}}
\put(2,0.4){\circle*{0.4}}
\put(1.8,0){\circle*{0.4}}
\put(2.8,0.2){\circle*{0.4}}
\put(3.2,0.6){\circle*{0.4}}
\put(2.7,1.1){\circle{0.4}}
\put(3,1.6){\circle*{0.4}}
\put(3.3,2){\circle*{0.4}}
\put(2.5,2.9){\circle*{0.4}}
\put(3.2,2.7){\circle*{0.4}}
\put(3.1,3.3){\circle*{0.4}}
\put(2.8,3.8){\circle*{0.4}}
\put(3.2,4.3){\circle*{0.4}}
\put(2.7,4.6){\circle*{0.4}}
\put(3,5.3){\circle*{0.4}}
\put(3.3,5.8){\circle*{0.4}}
\put(-9,5){(a)}
\thinlines
\put(-6,1.8){\circle*{0.54}}
\put(-5.9,2.1){\vector(1,2){0.9}}
\dashline{0.2}(-6.1,2.1)(-7,3.83)
\put(-7,3.83){\vector(-1,2){0}}
\put(-6,4.2){\circle{0.54}}
\put(-6.1,3.9){\vector(-1,-2){0.9}}
\dashline{0.2}(-5.9,3.9)(-5,2.17)
\put(-5,2.17){\vector(1,-2){0}}
\put(-7.6,2.2){\makebox(0,0){$P_1$}}
\put(-4.4,2.2){\makebox(0,0){$P_2$}}
\put(-7.6,3.8){\makebox(0,0){$P_2$}}
\put(-4.4,3.8){\makebox(0,0){$P_1$}}
\end{picture}
\capt{(a) For pedestrians with an opposite 
direction of motion it is advantageous
if both prefer either the right hand side or the left hand side when
trying to pass each other. Otherwise, they would have to stop in order
to avoid a collision. The probability $P_1$ of choosing the right hand side
is usually different from the probability $P_2=1-P_1$ 
of choosing the left hand side.\\
(b) Opposite directions of motion normally use separate lanes.
Avoidance maneuvers are indicated by 
arrows.\label{separation}}
\end{figure}
\begin{figure}[htbp]
\parbox[b]{7.8cm}{                                                   
\epsfxsize=7cm 
\centerline{\rotate[r]{\hbox{\epsffile[28 28 570                              
556]{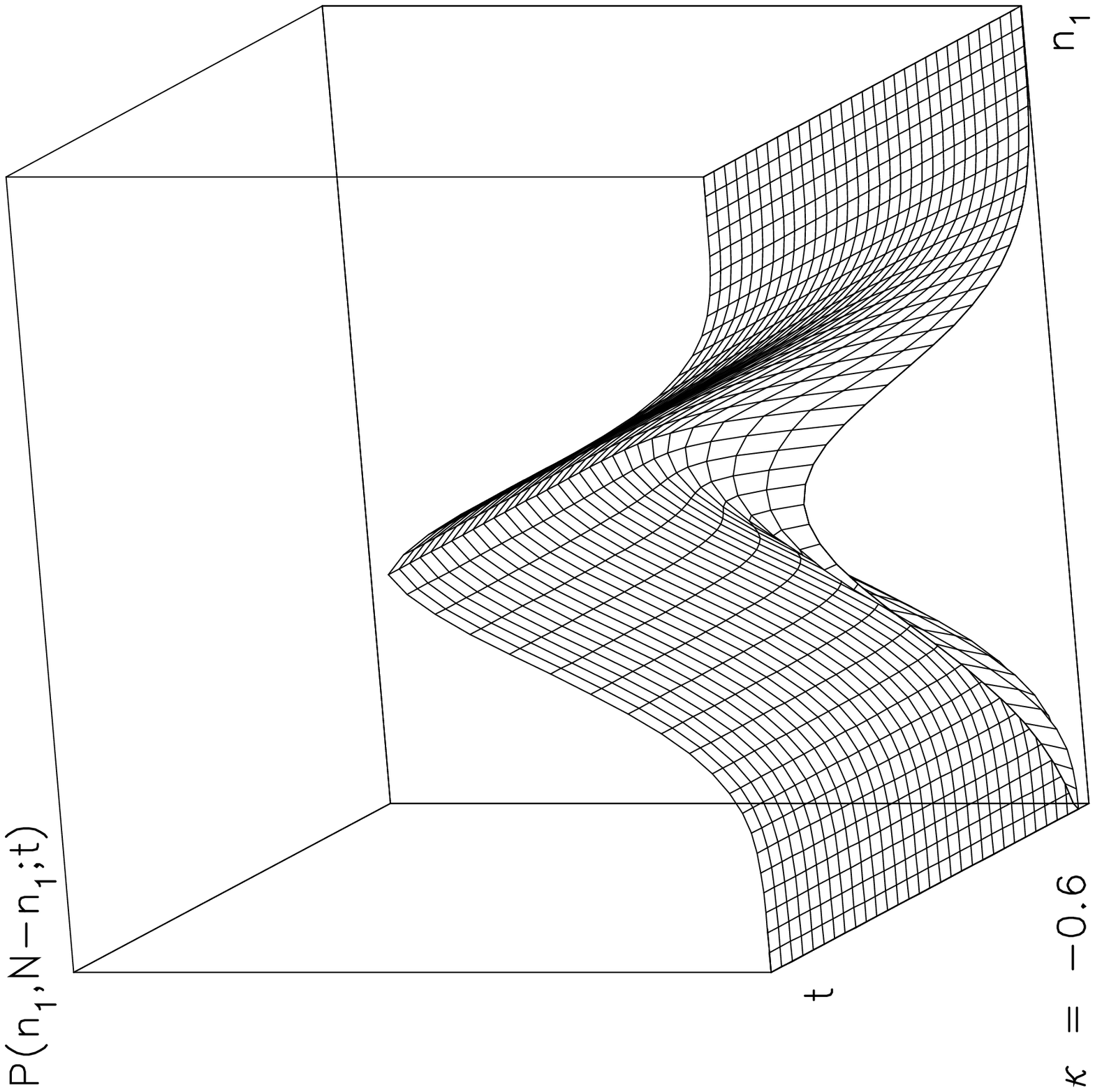}}}}
}\hfill
\parbox[b]{7.8cm}{
\epsfxsize=7cm 
\centerline{\rotate[r]{\hbox{\epsffile[28 28 570
556]{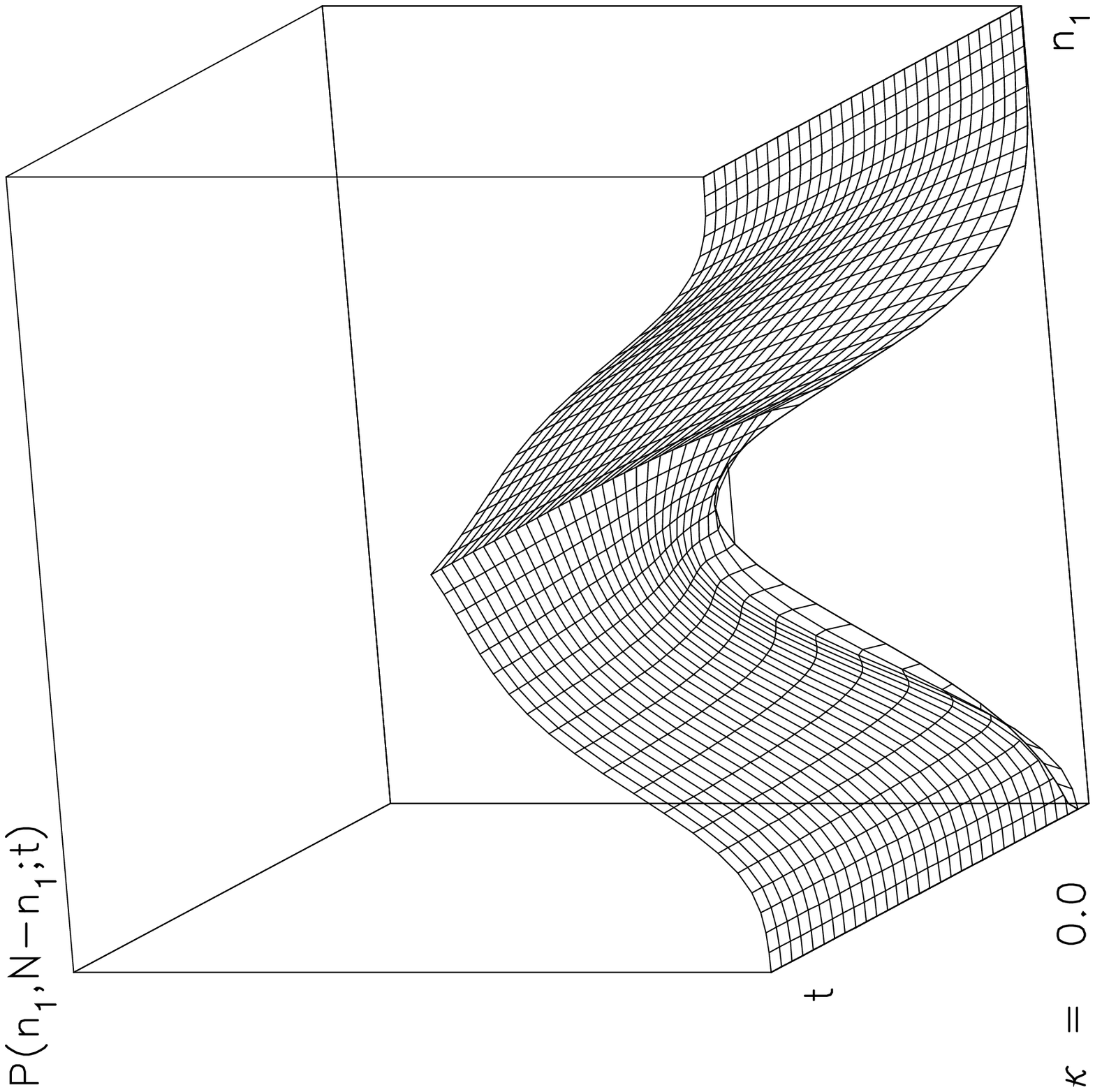}}}}
}                                              
%
\parbox[b]{7.8cm}{
\epsfxsize=7cm 
\centerline{\rotate[r]{\hbox{\epsffile[28 28 570
556]{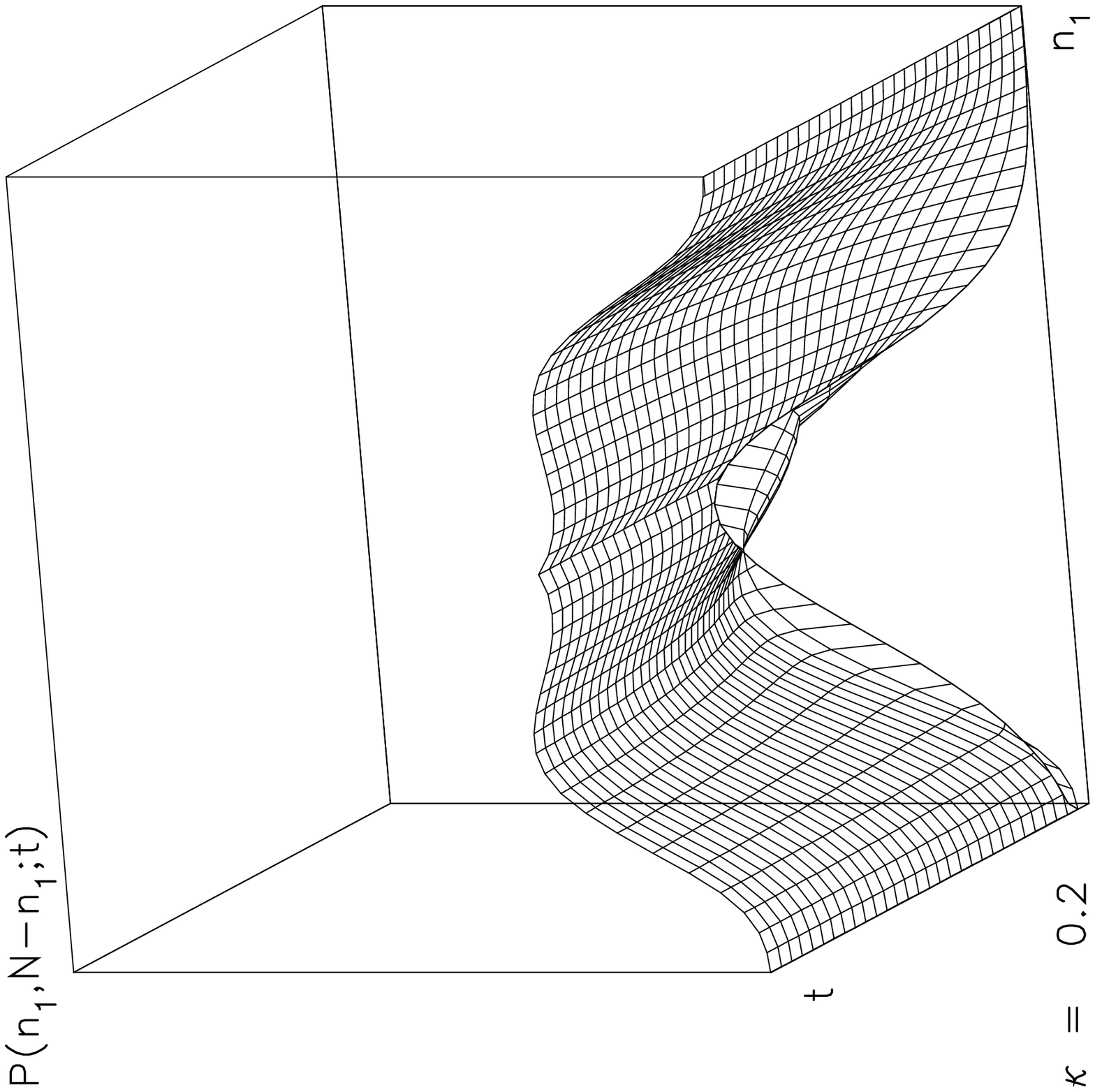}}}}
}\hfill
\parbox[b]{7.8cm}{
\epsfxsize=7cm 
\centerline{\rotate[r]{\hbox{\epsffile[28 28 570
556]{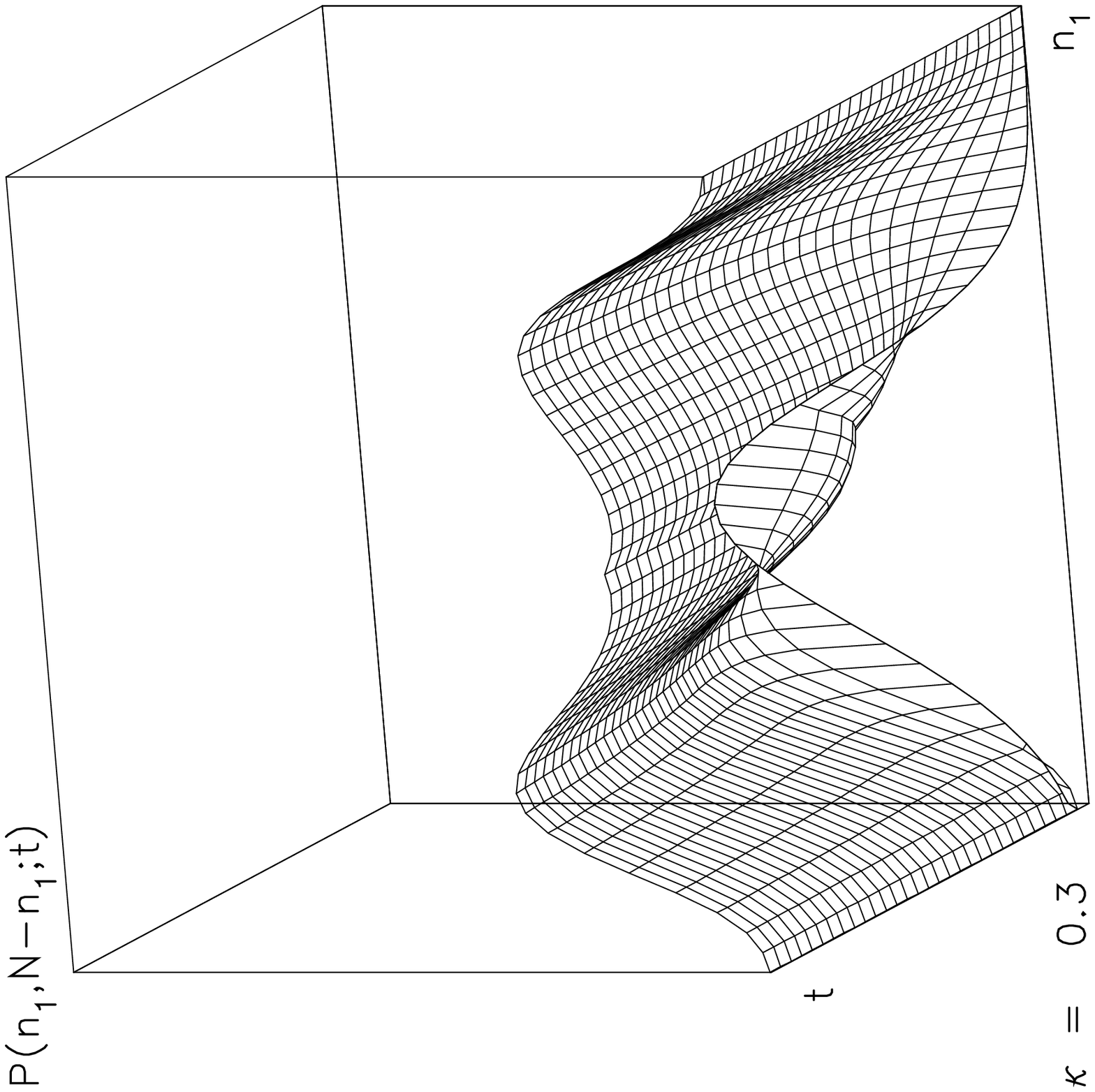}}}}
}
\parbox[b]{16cm}{\capt{Probability distribution $P(\vec{n},t)
\equiv P(n_1,N-n_1;t)$ of the socioconfiguration $\vec{n}$
for varying values of the control parameter $\kappa$. For $\kappa = 0$
a phase transition occurs: Whilst for $\kappa < 0$ both strategies
are used by about one half of the individuals, for $\kappa > 0$
very probably one of the strategies will be prefered after some time. 
That means, a behavioral convention
develops by social self-organization.\label{phasetr}}}
\end{figure}
\begin{figure}[htbp]
\epsfysize=12cm 
\centerline{\rotate[r]{\hbox{\epsffile[57 40 570 756]
{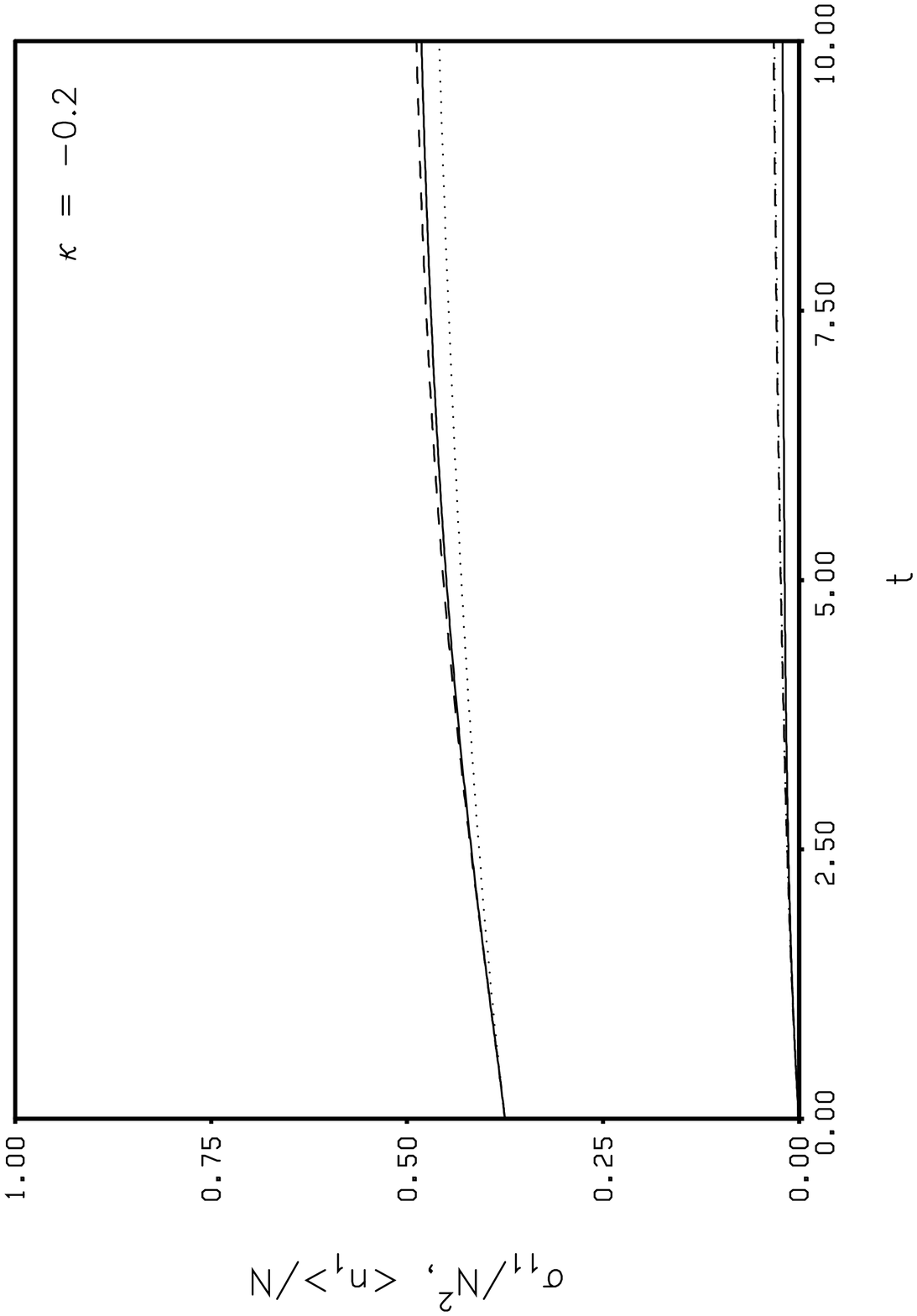}}}}
\capt{Exact (---), approximate ($\cdots$) and corrected (-- --) 
mean values (upper curves) and variances (lower curves)
for a {\em small} configurational distribution $P(\vec{n},t)$:
The simulation results for the approximate equations 
are acceptable, those for the corrected equations very well.
\label{below2}}
\end{figure}
%
\begin{figure}[htbp]
\epsfysize=12cm 
\centerline{\rotate[r]{\hbox{\epsffile[57 40 570 756]
{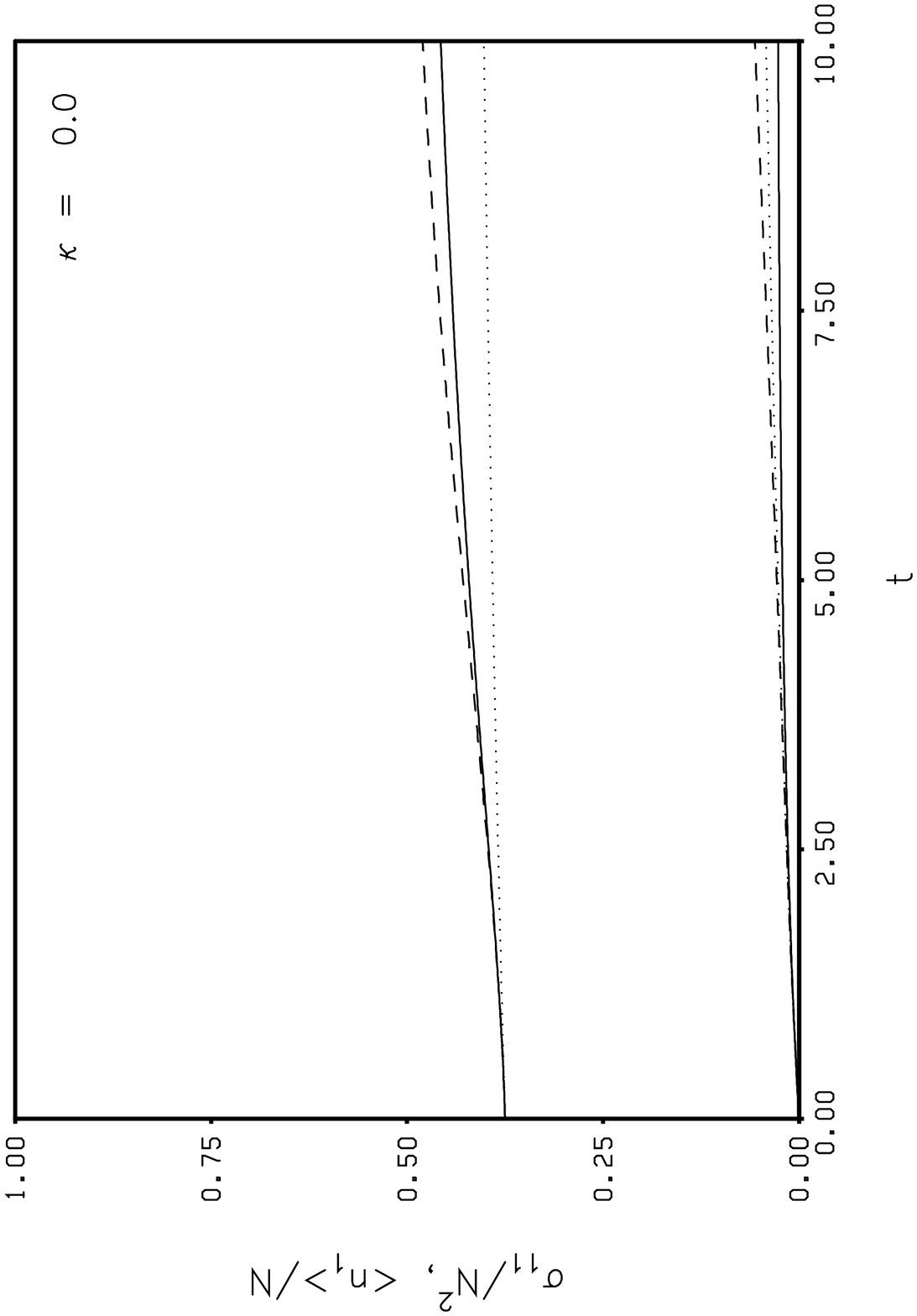}}}}
\capt{As Figure \ref{below2}, but for 
a {\em broad} configurational distribution:
The corrected equations still yield
useful results, whereas the approximate equations already fail
since the variances are not negligible.
\label{at2}}
\end{figure}\alphfig{valid}
%
%
%
\begin{figure}[htbp]
\epsfysize=12cm 
\centerline{\rotate[r]{\hbox{\epsffile[57 40 570 756]
{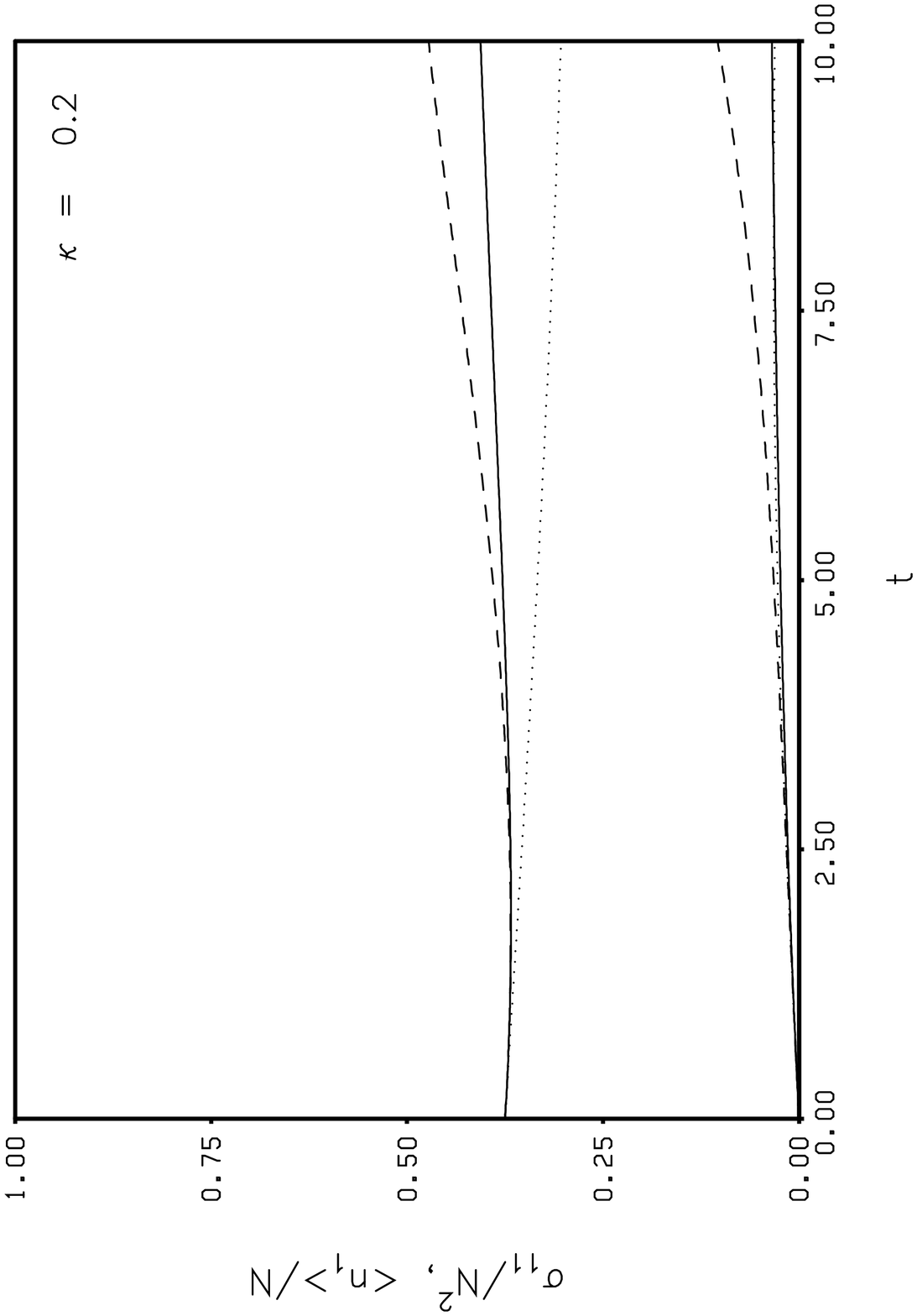}}}}
\capt{As Figure \ref{below2}, but for a {\em multimodal} configurational 
distribution: Not only the approximate but also
the corrected equations fail after a certain time interval. 
However, whereas the approximate mean value and variance become unreliable
already for $t > 1$, the corrected mean value and variance remain valid 
as long as $t\le 3$.\label{valid1}}
\end{figure}
\begin{figure}[htbp]
\epsfysize=12cm 
\centerline{\rotate[r]{\hbox{\epsffile[57 40 570 756]
{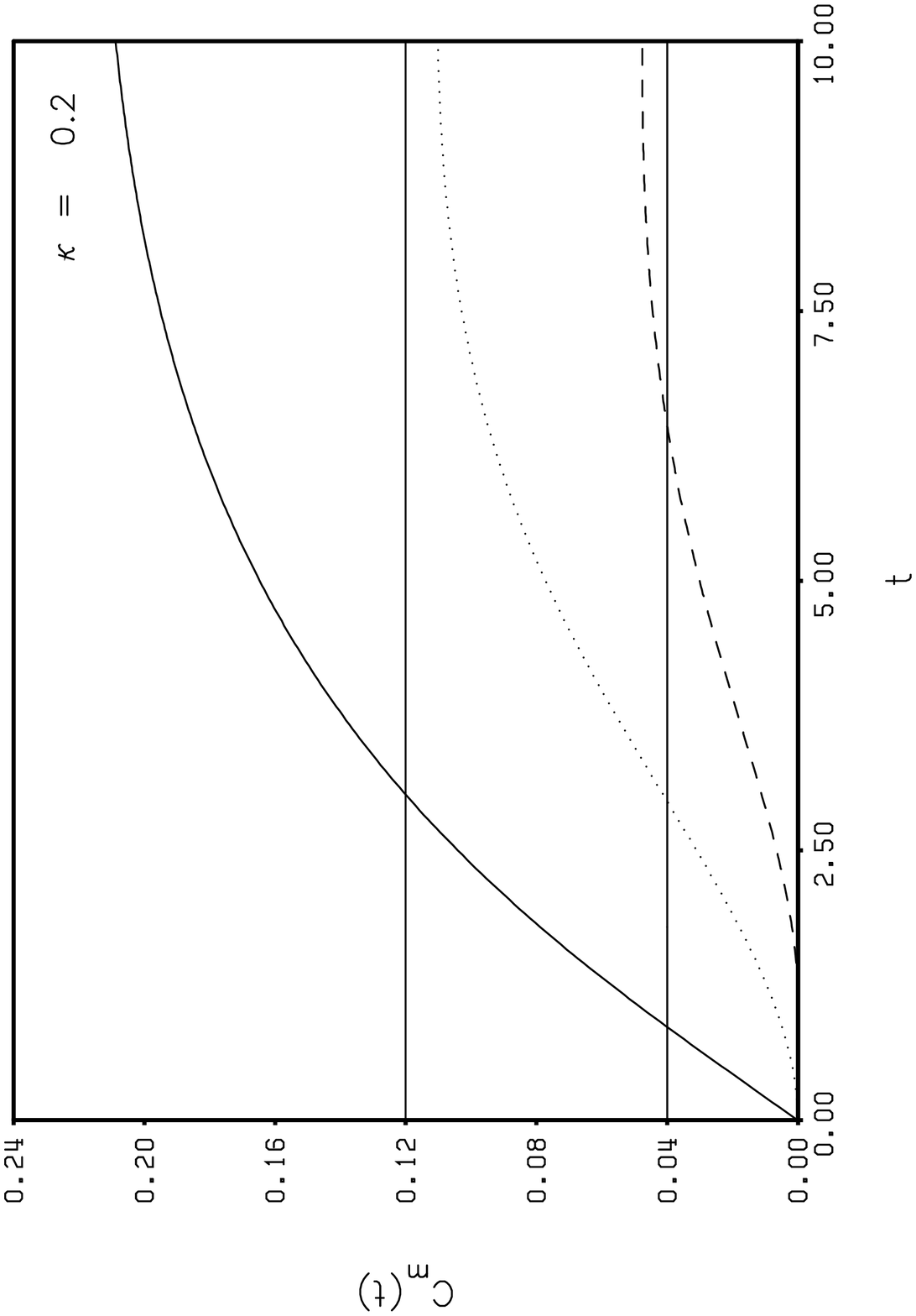}}}}
\capt{The relative central moments $C_m(t)$ are a criterium for the validity
of the approximate respectively the corrected mean value and covariance
equations: If $|C_2(t) |$ (---) exceeds the value 0.04, the approximate
equations fail. The corrected equations fail if $|C_3(t) |$ (-- --)
or $|C_4(t) |$ ($\cdots$) exceed the value 0.04. This is the case
if $|C_2(t) |$ becomes greater than 0.12 (indicating a
phase transition).\label{valid2}}
\end{figure}
\resetfig
\begin{figure}[htbp]
\parbox[b]{7.8cm}{                                                   
\epsfxsize=7cm 
\centerline{\rotate[r]{\hbox{\epsffile[28 28 570                              
556]{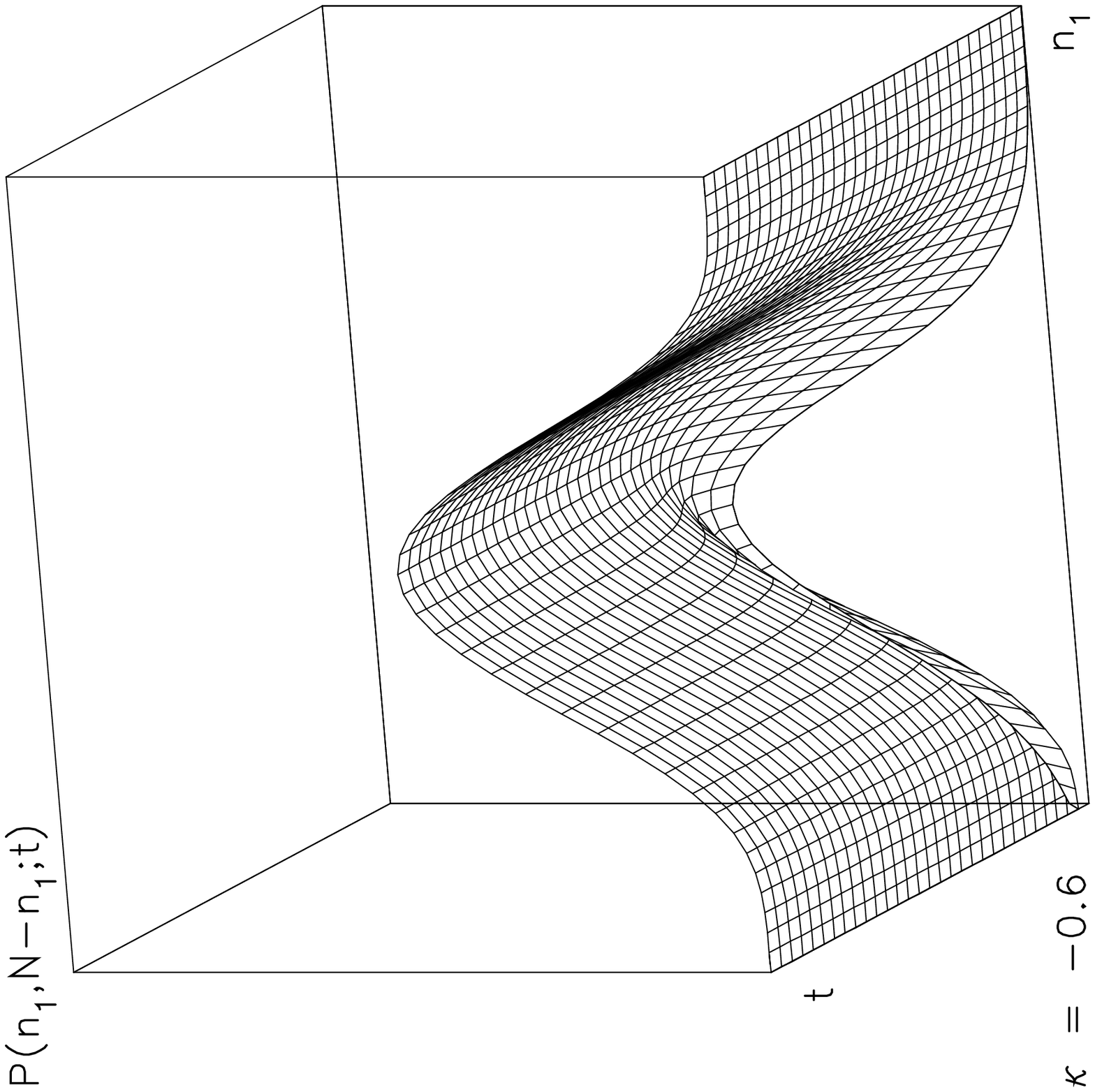}}}}
}\hfill
\parbox[b]{7.8cm}{
\epsfxsize=7cm 
\centerline{\rotate[r]{\hbox{\epsffile[28 28 570
556]{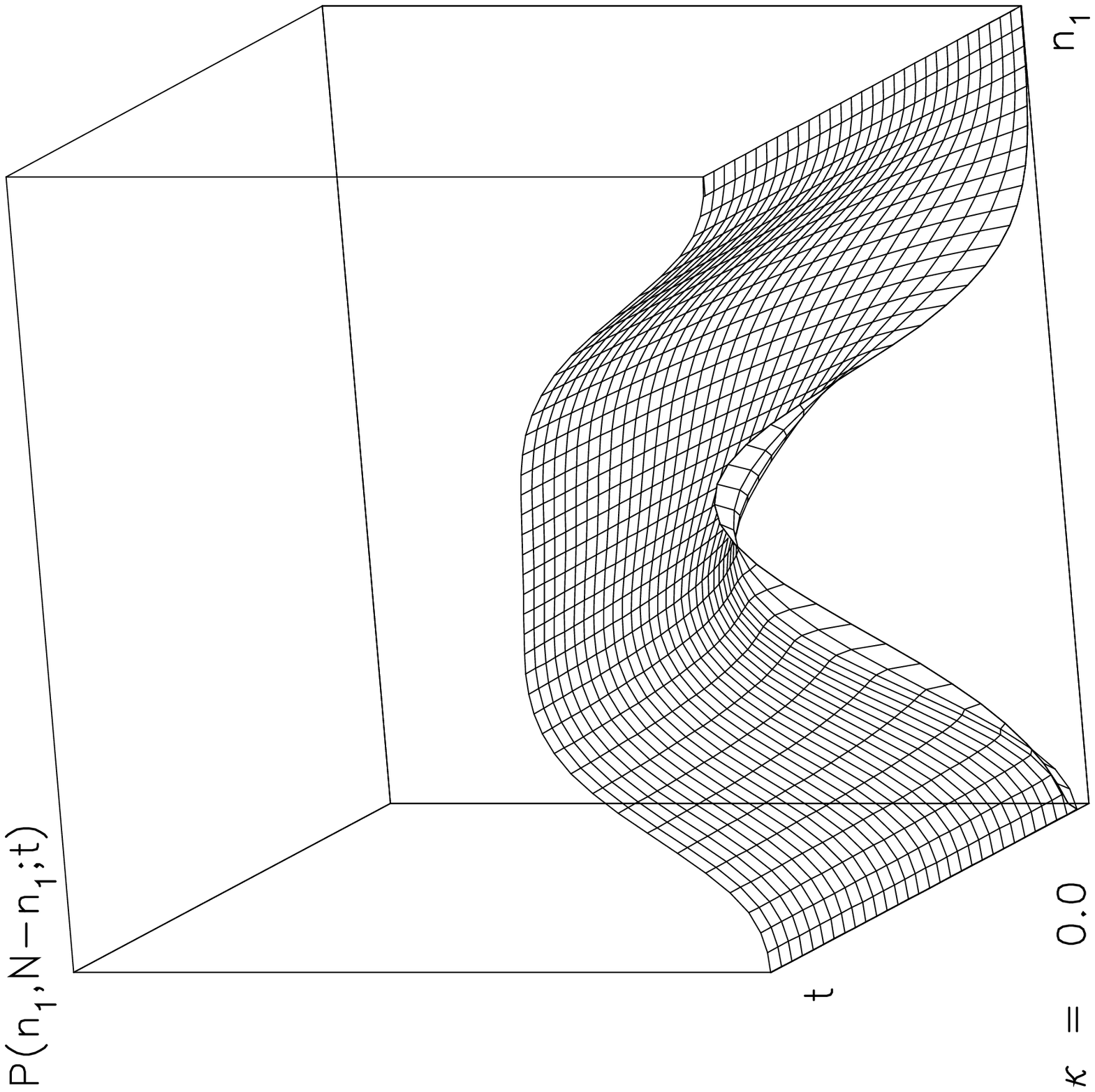}}}}
}                                              
%
\parbox[b]{7.8cm}{
\epsfxsize=7cm 
\centerline{\rotate[r]{\hbox{\epsffile[28 28 570
556]{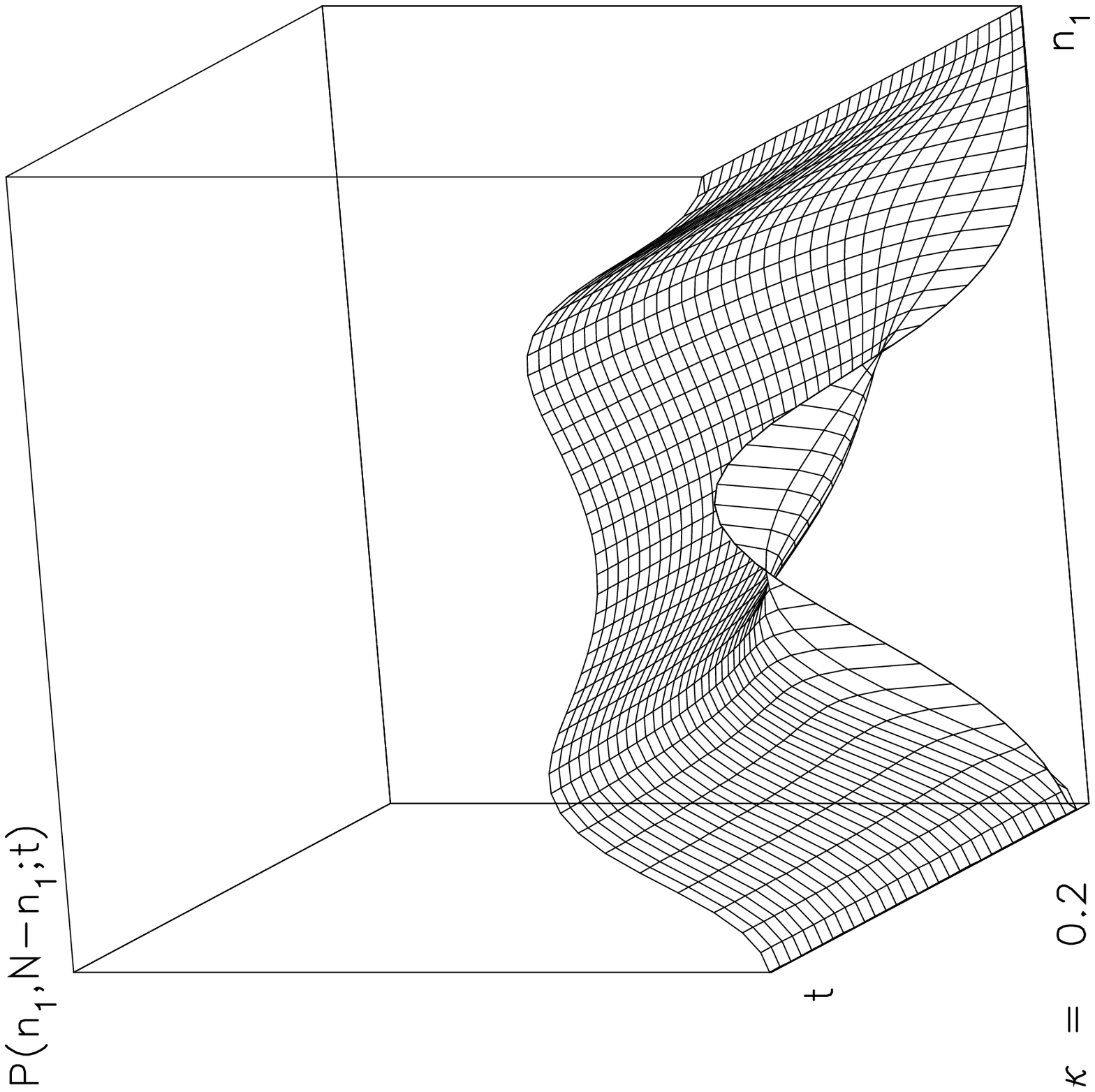}}}}
}\hfill
\parbox[b]{7.8cm}{
\epsfxsize=7cm 
\centerline{\rotate[r]{\hbox{\epsffile[28 28 570
556]{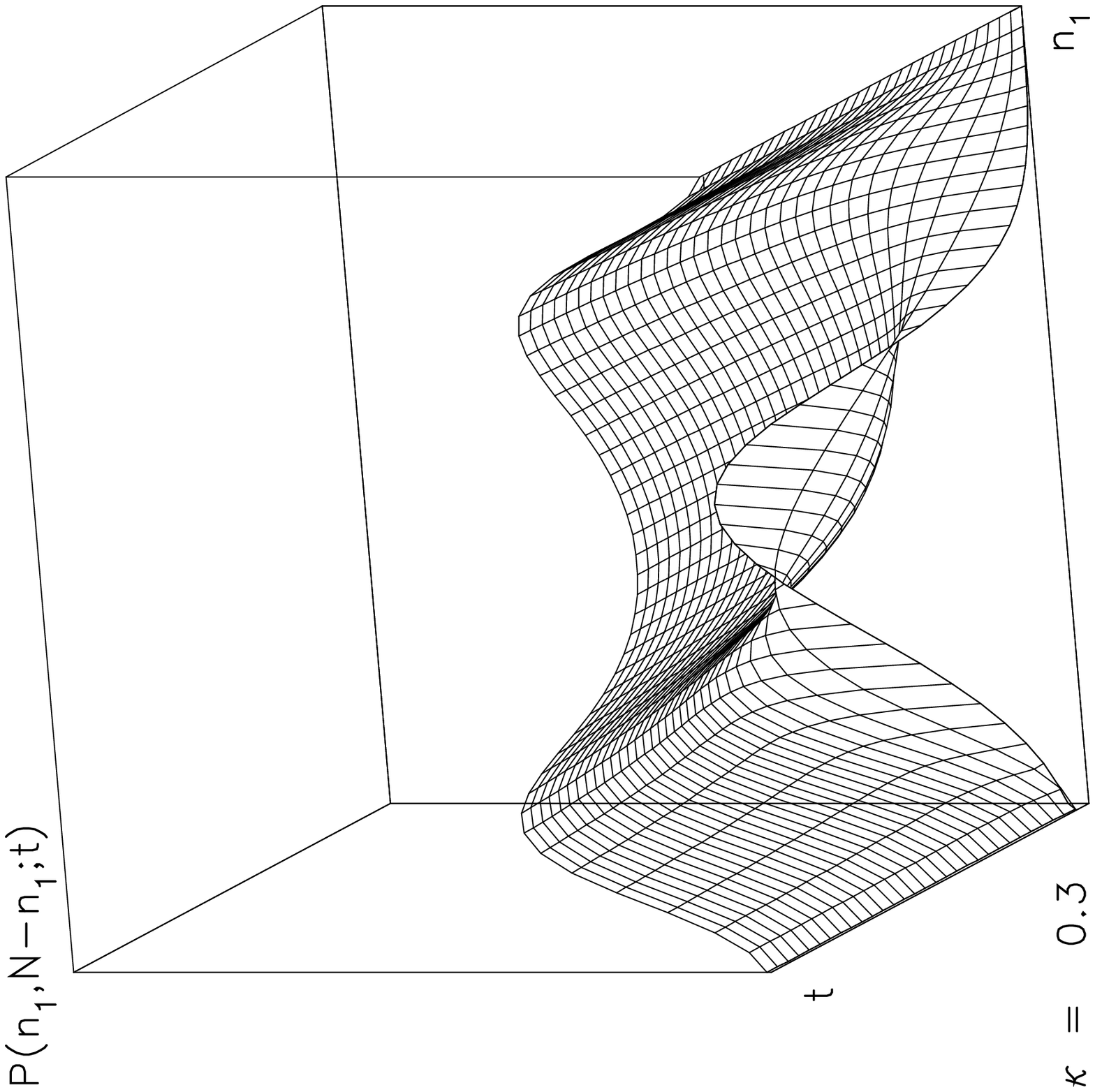}}}}
}
\parbox[b]{16cm}{\capt{Probability distribution $P(\vec{n},t)
\equiv P(n_1,N-n_1;t)$ of the socioconfiguration $\vec{n}$
according to the {\em modified} 
stochastic game dynamical equations. The results are
similar to those in Figure \ref{phasetr}. 
For $\kappa < 0$ both strategies
are used by about one half of the individuals, for $\kappa > 0$
very probably one of the strategies will be prefered after some time. 
Again, for $\kappa = 0$ a phase transition occurs.\label{crit}}}
\end{figure}

\clearpage
\begin{thebibliography}{20}\mbox{ }
\bibitem{Arthur1}Arthur, W. B.: 1988, `Competing Technologies: An Overview',
in: Dosi, R. et. al. (eds.), {\em Technical Change and Economic Theory},
Pinter Publishers, Londres and New York.
\bibitem{Arthur2}Arthur, W. B.: 1989, `Competing Technologies, Increasing
Returns, and Lock-In by Historical Events', {\em The Economic Journal} {\bf
99}, 116--131.
\bibitem{Axelrod}Axelrod, R.: 1984, {\em The Evolution of Cooperation},
 Basic Books, New York.
\bibitem{Boltzmann}Boltzmann, L.: 1964, {\em Lectures on Gas Theory},
 University of California, Berkeley.
\bibitem{Do}Domencich, T. A. and McFadden, D.: 1975, {\em Urban Travel
Demand. A Behavioral Ana\-ly\-sis}, North-Holland, Amsterdam, pp. 61-69.
\bibitem{Durlauf1}Durlauf, S.: 1989, `Locally Interacting Systems, Coordination
Failure, and the Long Run Behavior of Aggregate Activity', Working Paper No.
3719, National Bureau of Economic Research, Cambridge, MA.
\bibitem{Durlauf2}Durlauf, S.: 1991, `Nonergodic Economic Growth',
mimeo Stanford University. 
\bibitem{Eigen}Eigen, M.: 1971, `The Selforganization of Matter 
 and the Evolution
 of Biological Macromolecules', {\em Naturwissenschaften} {\bf 58}, 465.
\bibitem{Eig}Eigen, M. and Schuster, P.: 1979, {\em The Hypercycle},
 Springer, Berlin.
\bibitem{Feistel}Feistel, R. and Ebeling, W.: 1989, {\em Evolution of Complex
 Systems}, Kluwer Academic, Dordrecht. 
\bibitem{Fisher}Fisher, R. A.: 1930, {\em The Genetical Theory of Natural 
 Selection}, Oxford University, Oxford.
\bibitem{Fokker}Fokker, A. D.: 1914, {\em Annalen der Physik} {\bf 43},
810ff. 
\bibitem{Foellmer}F\"ollmer, H.: 1974, `Random Economics with Many Interacting
Agents', {\em Journal of Mathematical Economics} {\bf 1}, 51--62.
\bibitem{Gard}Gardiner, C. W.: 1983, {\em Handbook of Stochastic Methods},
Springer, Berlin.
\bibitem{Huber}Glance, N. S. and Huberman, B. A.: 1992, `Dynamics with
Expectations', {\em Physics Letters A} {\bf 165}, 432--440. 
\bibitem{Haag}Haag, G., Hilliges, M., and Teichmann, K.: 1993,
`Towards a Dynamic Disequilibrium Theory of Economy', in: Nijkamp, P. and
Reggiani, A., {\em Nonlinear Evolution of Spatial Economic Systems},
Springer, Berlin. 
\bibitem{Hak}Haken, H.: 1975, `Cooperative Phenomena in Systems Far from
Thermal Equilibrium and in Nonphysical Systems', {\em Reviews of Modern
Physics} {\bf 47}, 67--121.
\bibitem{Haken}Haken, H.: 1979, {\em Synergetics. An Introduction},
Springer, Berlin. 
\bibitem{Haken2}Haken, H.: 1983, {\em Advanced Synergetics},
Springer, Berlin.
\bibitem{Hauk}Hauk, M.: 1994, {\em Evolutorische \"Okonomik und private
Transaktionsmedien}, Lang, Frankfurt/Main.
\bibitem{Helbing}Helbing, D.: 1991, `A Mathematical Model for the Behavior of
Pedestrians', {\em Behavioral Science} {\bf 36}, 298-310.
\bibitem{Hel}Helbing, D.: 1992, {\em Stochastische Methoden, nichtlineare
Dynamik und quantitative Modelle sozialer Prozesse}, PhD thesis,
University of Stuttgart. Published 1993 by Shaker, Aachen. Corrected and
enlarged English edition: {\em Quantitative Sociodynamics. Stochastic
Methods and Models of Social Interaction Processes} published 1995 by Kluwer
Academic, Dordrecht. 
\bibitem{Helbinga}Helbing, D.: 1992a, `Interrelations between 
Stochastic Equations for Systems with Pair Interactions',
{\em Physica A} {\bf 181}, 29-52.
\bibitem{Helbingb}Helbing, D.: 1992b, `A Mathematical Model for Attitude 
Formation by Pair Interactions', {\em Behavioral Science} {\bf 37}, 190-214.
\bibitem{Helb}Helbing, D.: 1994, `A Mathematical Model for the Behavior
of Individuals in a Social Field', {\em Journal of Mathematical Sociology}
{\bf 19}, 189--219.
\bibitem{Hofb}Hofbauer, J., Schuster, P., and Sigmund, K.: 1979,
`A Note on Evolutionarily Stable Strategies and Game Dynamics',
{\em J. theor. Biology} {\bf 81}, 609-612.
\bibitem{Ho}Hofbauer, J., Schuster, P., Sigmund, K., and Wolff, R.: 1980,
`Dynamical Systems under Constant Organization', {\em J. Appl. Math.}
{\bf 38}, 282-304.
\bibitem{Hofbauer}Hofbauer, J. and Sigmund, K.: 1988, {\em The Theory of 
Evolution and Dynamical Systems}, Cambridge University, Cambridge.
\bibitem{Ising}Ising, E.: 1925, {\em Zeitschrift f\"ur Physik} {\bf 31},
253ff.
\bibitem{Kram}Kramers, H. A.: 1940, {\em Physica} {\bf 7}, 284ff. 
\bibitem{Lan}Langevin, P.: 1908, {\em Comptes. Rendues} {\bf 146}, 530ff.
\bibitem{Luce}Luce, R. D. and Raiffa, H.: 1957, {\em Games and Decisions},
Wiley, New York.
\bibitem{Lu}Luce, R. D.: 1959, {\em Individual Choice Behavior}, Wiley,
New York, Chap. 2.A: `Fechner's Problem'.
\bibitem{Moyal}Moyal, J. E.: 1949, {\em J. R. Stat. Soc.} {\bf 11}, 151--210.
\bibitem{Neumann}von Neumann, J. and Morgenstern, O.: 1944, {\em Theory of 
Games and Economic Behavior}, Princeton University, Princeton.
\bibitem{Prigogine2}Nicolis, G. and Prigogine, I.: 1977, {\em Self-Organization
in Nonequilibrium Systems}, Wiley, New York.
\bibitem{Orlean1}Orl\'{e}an, A.: 1992, `Contagion des Opinions et
Fonctionnement des March\'{e}s Financiers', {\em Revue \'{E}conomique}
{\bf 43}, 685--698.
\bibitem{Orlean2}Orl\'{e}an, A.: 1993, Dezentralized Collective Learning and
Imitation: A Quantitative Approach', mimeo CREA. 
\bibitem{Orlean3}Orl\'{e}an, A. and Robin, J.-M.: 1992, `Variability of
Opinions and Speculative Dynamics on the Market of a Storable Goods', mimeo
CREA. 
\bibitem{Pauli}Pauli, H.: 1928, in: Debye, P. (ed.),
{\em Probleme der Modernen Physik}, Hirzel, Leipzig.
\bibitem{Planck}Planck, M.: 1917, in {\em Sitzungsber. Preuss. Akad. Wiss.},
pp. 324ff.
\bibitem{Prigogine1}Prigogine, I.: 1976, `Order through Fluctuation:
Self-Organization and Social System', in: Jantsch, E. and Waddington, C. H.
(eds.), {\em Evolution and Consciousness. Human Systems in Transition},
Addison-Wesley, Reading, MA.
\bibitem{Rap}Rapoport, A. and Chammah, A. M.: 1965, {\em Prisoner's Dilemma.
A Study in Conflict and Cooperation}, University of Michigan Press, Ann Arbor.
\bibitem{Schn}Schnabl, W., Stadler, P. F., Forst, C., and Schuster, P.: 1991,
`Full Characterization of a Strange Attractor', {\em Physica D} {\bf 48},
65-90.
\bibitem{Schust}Schuster, P., Sigmund, K., Hofbauer, J., and Wolff, R.: 1981, 
`Selfregulation of Behavior in Animal Societies', {\em Biological 
Cybernetics} {\bf 40}, 1-25.
\bibitem{Strat}Stratonovich, R. L.: 1963, 1967, {\em Topics in the Theory of
Random Noise}, Vols. 1 \& 2, Gordon and Breach, New York.
\bibitem{Tay}Taylor, P. and Jonker, L.: 1978, `Evolutionarily Stable
Strategies and Game Dynamics', {\em Math. Biosciences} {\bf 40}, 145-156.
\bibitem{Topol}Topol, R.: 1991, `Bubbles and Volatility of Stock Prices:
Effect of Mimetic Contagion', {\em The Economic Journal} {\bf 101}, 786--800.
\bibitem{W}Weidlich, W.: 1971, `The Statistical Description of Polarization
Phenomena in Society', {\em Br. J. Math. Stat. Psychol.} {\bf 24}, 51ff.
\bibitem{We}Weidlich, W.: 1972, `The Use of Statistical Models in Sociology',
{\em Collective Phenomena} {\bf 1}, 51--59.
\bibitem{Wei}Weidlich, W.: 1991, `Physics and Social Science---The Approach
of Synergetics', {\em Physics Reports} {\bf 204}, 1-163.
\bibitem{WeiBrau}Weidlich, W. and Braun, M.: 1992, `The Master Equation
Approach to Nonlinear Economics', {\em Journal of Evolutionary Economics} {\bf
2}, 233--265.
\bibitem{Weid1}Weidlich, W. and Haag, G.: 1983, {\em Concepts and Models of a
Quantitative Sociology. The Dynamics of Interacting Populations},
Springer, Berlin.
\bibitem{Zee}Zeeman, E. C.: 1980, `Population Dynamics from Game Theory',
in: {\em Global Theory of Dynamical Systems}, Lecture Notes in Mathematics
{\bf 819}.\\[1.2cm]
\end{thebibliography}
\end{document}